\documentclass[10pt, letterpaper]{article}
\usepackage{mathrsfs}
\usepackage{epic, eepic}
\usepackage{subfigure}
\usepackage{amsfonts, amssymb, amsmath, graphicx, epsfig, lscape, capt-of}
\usepackage{url}
\usepackage{hyperref}
\usepackage{ifpdf}
\usepackage[boxed]{algorithm2e}
\usepackage{array}
\usepackage{booktabs}
\usepackage{multirow}
\usepackage{xcolor}
\usepackage{adjustbox}
\usepackage{longtable}
\usepackage{booktabs}

\def\inst#1{$^{#1}$}

\makeatletter

\begin{document}%

%%%%%%%%%%%%%%%%%%%%%%

\title{Anxiety for the pandemic and trust in financial markets\thanks{The authors sincerely thank Prof. Anna Maria D'Arcangelis for her contributions.}}

\author{%
Roy Cerqueti\inst{1,2}\footnote{Corresponding author.} \and Valerio Ficcadenti\inst{1}}
\date{}

\maketitle

\begin{center}
{\footnotesize

\vspace{0.3cm} \inst{1} School of Business\\
London South Bank University\\
London, SE1 0AA, UK\\
\texttt{ficcadv2@lsbu.ac.uk (V. Ficcadenti); cerquetr@lsbu.ac.uk (R. Cerqueti)}\\
\vspace{0.3cm} \inst{2} Department of Social and Economic Sciences \\
Sapienza University of Rome \\
Rome,  I-00185, Italy\\
\texttt{roy.cerqueti@uniroma1.it}}

\end{center}

\begin{abstract}
{The COVID-19 pandemic has generated disruptive changes in many fields. Here we focus on the relationship between the anxiety felt by people during the pandemic and the trust in the future performance of financial markets. Precisely, we move from the idea that the volume of Google searches about ``coronavirus'' can be considered as a proxy of the anxiety and, jointly with the stock index prices, can be used to produce mood indicators -- in terms of pessimism and optimism -- at country level. We analyse the ``very high human developed countries'' according to the Human Development Index plus China and their respective main stock market indexes. Namely, we propose both a temporal and a global measure of pessimism and optimism and provide accordingly a classification of indexes and countries. \\
The results show the existence of different clusters of countries and markets in terms of pessimism and optimism. Moreover, specific regimes along the time emerge, with an increasing optimism spreading during the mid of June 2020. Furthermore, countries with different government responses to the pandemic have experienced different levels of mood indicators, so that countries with less strict lockdown had a higher level of optimism.}
\end{abstract}

%%%%%%%%%%%%%%%%%%%%%%
\section{Introduction}

The world is experiencing the rapid and dramatic widespread of COVID-19 \cite{l20,z20} -- a pandemic generated by a coronavirus -- with millions of infected and a large number of deaths. Beyond the sanitary aspects of such an infectious disease, one of the main concerns experienced by the community regards the economic impact of the measures taken for contrasting the virus.

The individuals' behaviours are at the core of the interest of many scientific studies given that, those behaviours are the critical variables to understand the perspective of many economic activities. Several businesses require the physical interactions among the involved actors -- and such interactions have been reduced by the lockdown policies and by the natural attitude of people avoiding possible sources of contagion -- while virtual connections allow another set of economic relevant activities, such as investing in financial markets.
Interestingly, \cite{go90} provides a brief discussion on the reactions of the financial markets to rare catastrophic events of non-financial nature. The author points the attention of the readers to the plausible parallelisms between pandemics and natural disasters, terrorist attacks and even nuclear conflict. Less recently, \cite{ka10} elaborate on how aviation disasters can generate a decline in stock market prices. In general, empirical evidence prove that prices collapse in concomitance to rare and unexpected disasters (see, e.g. \cite{b06,ga12,go12}). On the same line, but in a broader perspective, several authoritative studies highlight that anxiety and negative mood might increase investors' risk aversion, hence leading to the collapse of stock prices (see, e.g. \cite{a90,ka00,ka03,c13}).

Therefore, the financial distress we are observing in the international stock markets can be reasonably interpreted through the anxiety of people, whose worries for the pandemic disease affect the expectation of financial markets future performance.

This paper enters this debate. Specifically, it explores how the
anxiety for COVID-19 mirrors the strategies of investing/disinvesting money in financial markets. In particular, we discuss the relationship between anxiety for COVID-19 and the view of financial markets, with the particular aim of investigating optimism and pessimism. The analysis is carried out by dealing with the country-level moods. We explore the relationship mentioned above for a large set of countries, to derive the different behaviours of the populations. \cite{f20} and \cite{b20} are remarkably relevant for contextualizing our study. The authors discuss the economic anxiety stemmed from coronavirus. \cite{b20} conducts a survey study of over 500 US consumers and shows that the serious concern about coronavirus implications leads to pessimistic expectations about macroeconomic turnaround via deterioration of the economic fundamentals. \cite{f20} complement \cite{b20}'s perspective by including also the time dimension and the causal effect of the pandemic on the increased economic anxiety. The methodological ground of \cite{f20} lies on the meaningfulness of Google Trends data, which are assumed to give in-depth information on the development of the anxiety in the specific context of the economic outcomes. We adopt \cite{f20}'s view and hypothesize that anxiety for COVID-19 is proxied by the irrepressible persistence of related Google searches. In so doing, we also follow \cite{m20}, where a survey study over a large number of respondents confirms that media exposure and online searches are good predictors of the increasing fear of coronavirus (in this, see also the review paper by \cite{ga20}).

In details, we collect and compare two different datasets over the same reference period which goes from January $6^{th}$, 2020 to the June $19^{th}$, 2020. By one side, we consider the daily Google Trends data. Specifically, we examine for the searches volumes of the word \textit{``coronavirus''} along with its translations for different countries and respective most spoken languages. Data retrieved at a country level allows for sounding out similarities and discrepancies in the searching for information practised by users in need of awareness. In our approach, such a compulsive searching is intended as a proxy for the anxiety generated by the pandemic. On the other side, we consider the daily levels of the main stock indexes, which include companies related to the considered countries. The source of financial data is Datastream.
In order to have a reliable and consistent dataset, countries are chosen by using the Human Development Index (HDI) embraced by the United Nations Development Programme (UNDP) in the Human Development Report Office to rank countries on the basis of their human development. Specifically, we select the areas having an HDI index greater than 0.8 calculated with the 2018 information. The choice of 0.8 as a threshold is appropriate because all the countries having at least that level can be considered as ``very high human developed countries''. It ensures a good enough level of connections between socio-financial entities within the countries, namely it guarantees the incorporation of the necessary links between citizens' cognitions of the problems, ability to get informed about them and financial strategies designer presence. To such a list of nations, we add China, which is ranked below the 0.8 thresholds -- specifically, 0.75. We reasonably do so because China is central in the phenomenon under investigation. Moreover, the countries without data on stock exchanges in our source -- which is Datastream -- have been obviously excluded from the list.

We move our steps from \cite{f20} in two main respects: first, the quoted paper deals with topics in Google Trends, and we deal with one crucial word. In so doing, we have a translation task to face -- as also acknowledged by \cite{f20}. Nevertheless, the use of one word allows to obtain intuitive results and is far from being restrictive in our context. Indeed, a preliminary inspection of the Google Trends data shows that the considered word is the most relevant trend related to the current pandemic; second, the quoted paper derives information about economic anxiety directly by Google Trends. Differently, we here start from the idea that the anxiety is manifested through the Google searches of the word \textit{``coronavirus''} (and its translations), but then we move to the real performance of the financial markets, to assess the links with the trust on them.

Some distance measures between time series have been suitably introduced to offer a wide perspective on the connections between the considered data.
We consider concepts of distance focusing on specific dates and offering global information on the entire reference period. All the proposed measures range in the unitary interval $[0,1]$, hence letting the comparative analysis of different countries be possible.

Several interesting results emerge. Countries and markets can be properly clustered in terms of their mood during the pandemic period. Regularities and deviations at individual week level can be also identified. Moreover, the analysis of the daily variations of the levels of anxiety and trust in financial markets gives insights on countries behaviors in the overall period. A general trend of pessimism is concentrated in early and mid March, when the lockdown have been adopted by many countries and the international community started to reckon the severity of the problem. A focus on some noticeable cases of hard and weak lockdown policies has been also presented. In this respect, countries with a stricter lockdown have a more persistent and higher level of pessimism.

The rest of the paper is organized as follows. Section \ref{data} presents the employed dataset, by providing also details on the data collection procedure. Section \ref{method} illustrates the methodological devices used for the study. Section
\ref{results} outlines and discusses the results of the analysis. Last section concludes.
%and traces the lines of future research themes.

\section{Data}
\label{data}

We now present the employed data. As we will see in detail below, the considered dataset is associated with the Google Trends and to the financial markets at country level. As a premise, we have to say that data on financial markets are not always available; moreover, the access to the web is not a reliable issue is some realities. Thus, we focus on a set of countries whose data are unbiased and reliable. At this aim -- and for providing a consistent analysis -- we have used the Human Development Index (HDI) adopted by United Nations Development Programme (UNDP)'s Human Development Report Office. HDI is a composite index made of factors like life expectancy, education, per capita income indicators and other relevant factors whose details are recollected in \cite{ul1995reflections} by Mahbub ul Haq, one of the two designers of the index. HDI is used to rank countries on the basis of human development. We take all the countries defined as ``very high human developed countries'', namely those having an HDI index greater than 0.8. The selection is based on data from 2018, Table 1 of \cite{HDR2019}. China is added to the considered countries -- even if the HDI of China is 0.75 -- because of its centrality in the COVID-19 propagation; the first known human infections were in China.

The respective most used language is associated with each of these countries.
Then, by means of Google Translate, the word \textit{``coronavirus''} is translated from English to each of those languages. In so doing, we obtain the translations reported in Table \ref{tab:1}.

The translated terms are employed to download the web searches indicator from Google Trends. Namely, for each country, one looks for the searches of the respective \textit{``coronavirus''} translations. In this way, the magnitude of the searches by country is obtained employing the words translated in the country most common language. The period investigated goes from January $6^{th}$, 2020 to June $19^{th}$, 2020.\\

\begin{figure}[!htb]
\includegraphics[width=\textwidth]{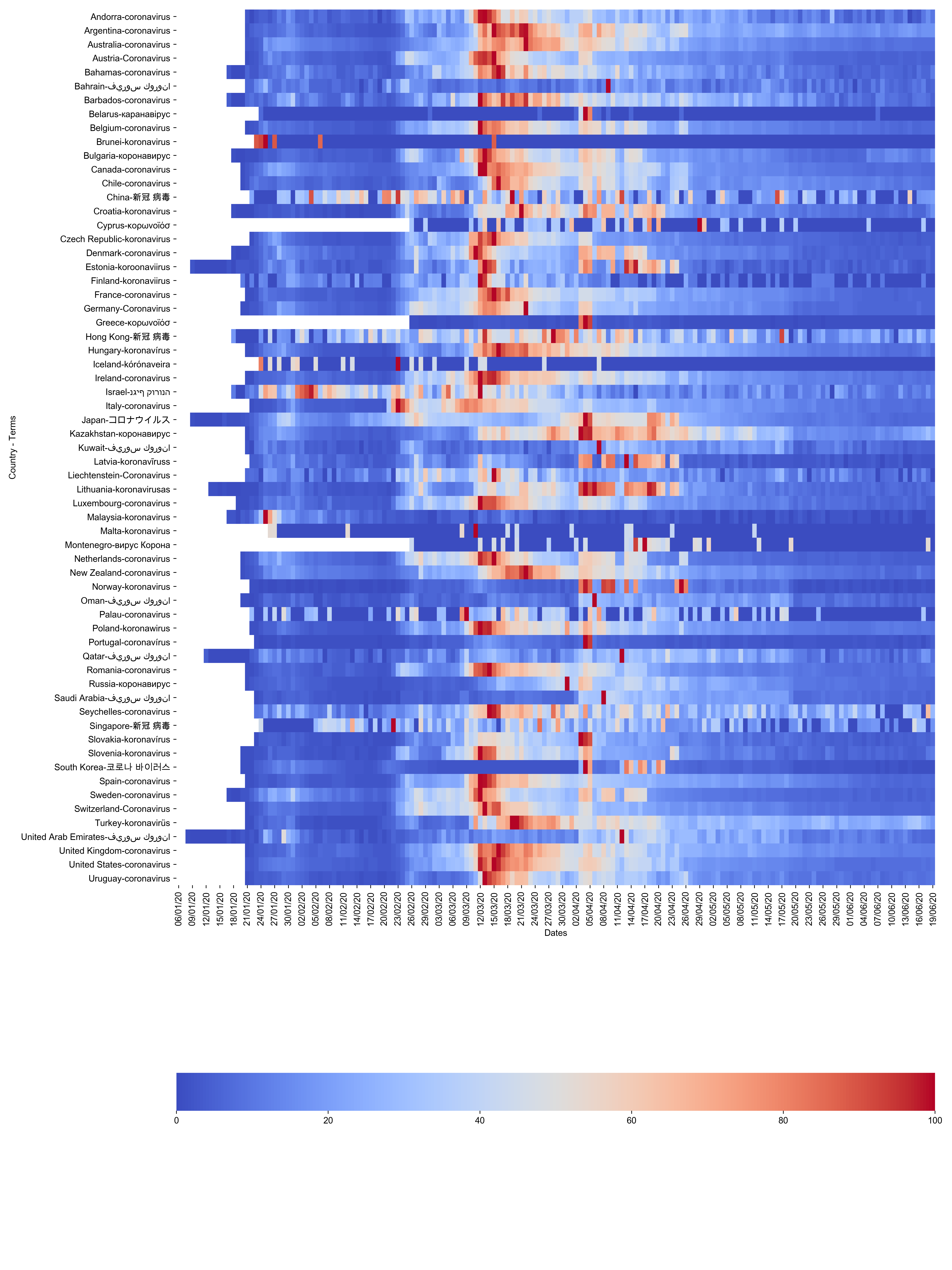}
\caption{Heatmap representation of the Google search indicators of the word \textit{``coronavirus''} and its translations in the respective most spoken language for each country. The indents give a clear view of the beginning of the interest in COVID-19 for each country.}
\label{fig:1}
\end{figure}

At the end of this process, one gets a matrix of time series regarding 63 countries. In our analysis, we are interested in examining the time series of the searches from the first day in which a relevant volume of researches is recorded in each country -- i.e., in the first day in which Google Trends offers a nonnull value -- for the respective translated terms. See columns one, two and three of Table \ref{tab:1} and Figure \ref{fig:1} to have an idea of the main trends in the data. The most evident point regards the high volume of searches occurred during the same days around mid-March 2020.

We associate at least one stock market index with each country of the list mentioned above. Per each index, the closing prices are downloaded from Thomson Reuters Datastream. The time span is defined by the same criterion adopted in collecting the Google Trends data (see Table \ref{tab:2} and Figure \ref{fig:2}), so that one has the same time span. Andorra, Bahamas, Barbados, Belarus, Brunei, Liechtenstein, Palau, Seychelles and Uruguay do not have a stock market index of reference in our data source, so we exclude them. The final list of considered countries contains 54 elements. Furthermore, we align the Google Trends data and the financial data so that, for each day in which prices are recorded, the volume of web searches can be used in the analysis. Consequently, because the financial markets are closed during non-trading days, Google Trends data is reduced accordingly. As a reference for the number of observation, one can look at column ``N. Obs.'' in Table \ref{tab:2}.

\begin{table}[!htb]
\centering
\includegraphics[scale = 0.68]{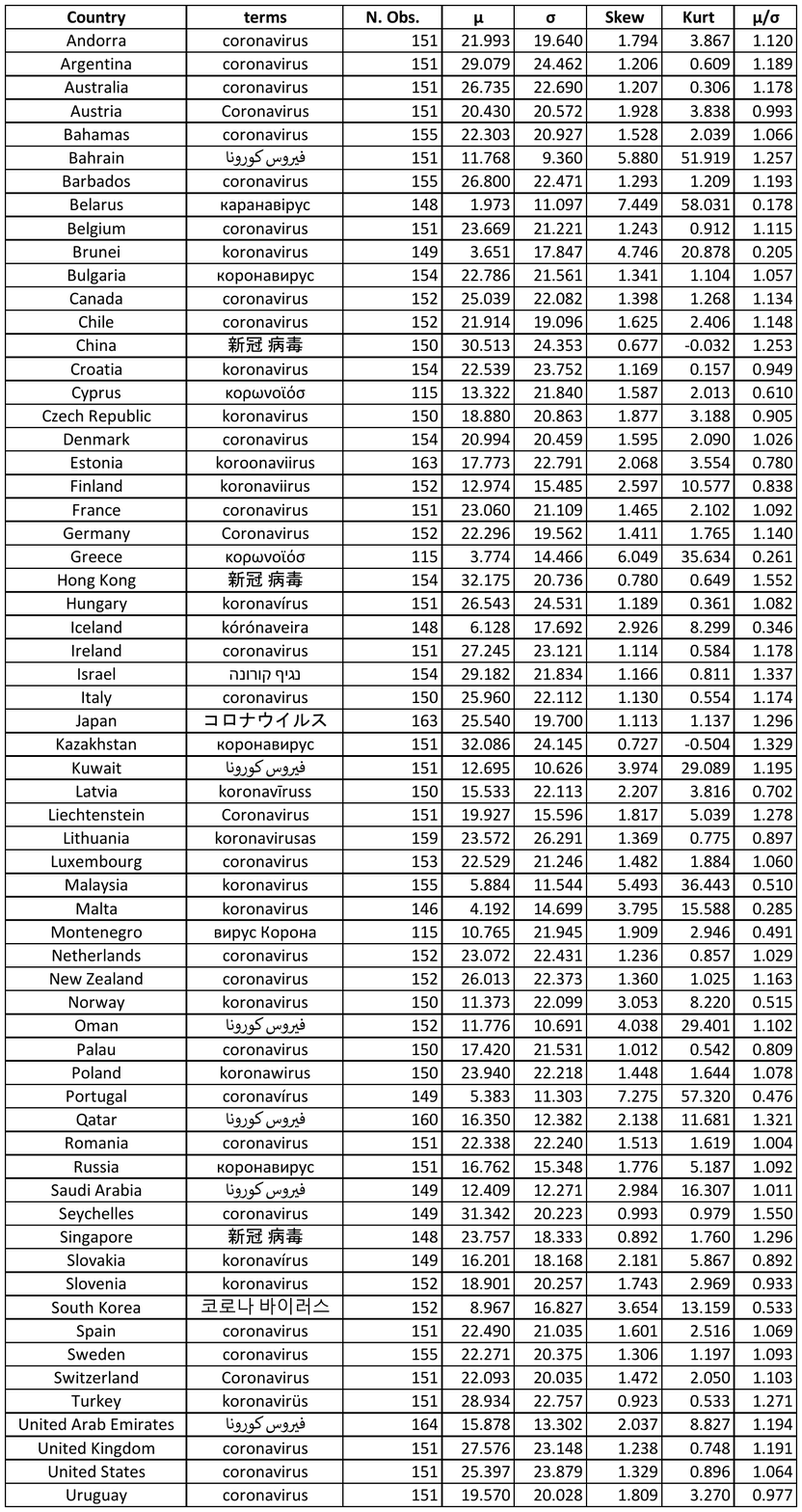}
\caption{Google Trends data. The table contains country name, translation of \textit{``coronavirus''} from English to the most used language in the respective country and statistical summary of the related  time series. The different number of observations depends on the first date in which a positive value for the search volumes is recorded.}
\label{tab:1}
\end{table}

\begin{table}[!htb]
\centering
\includegraphics[scale = 0.75]{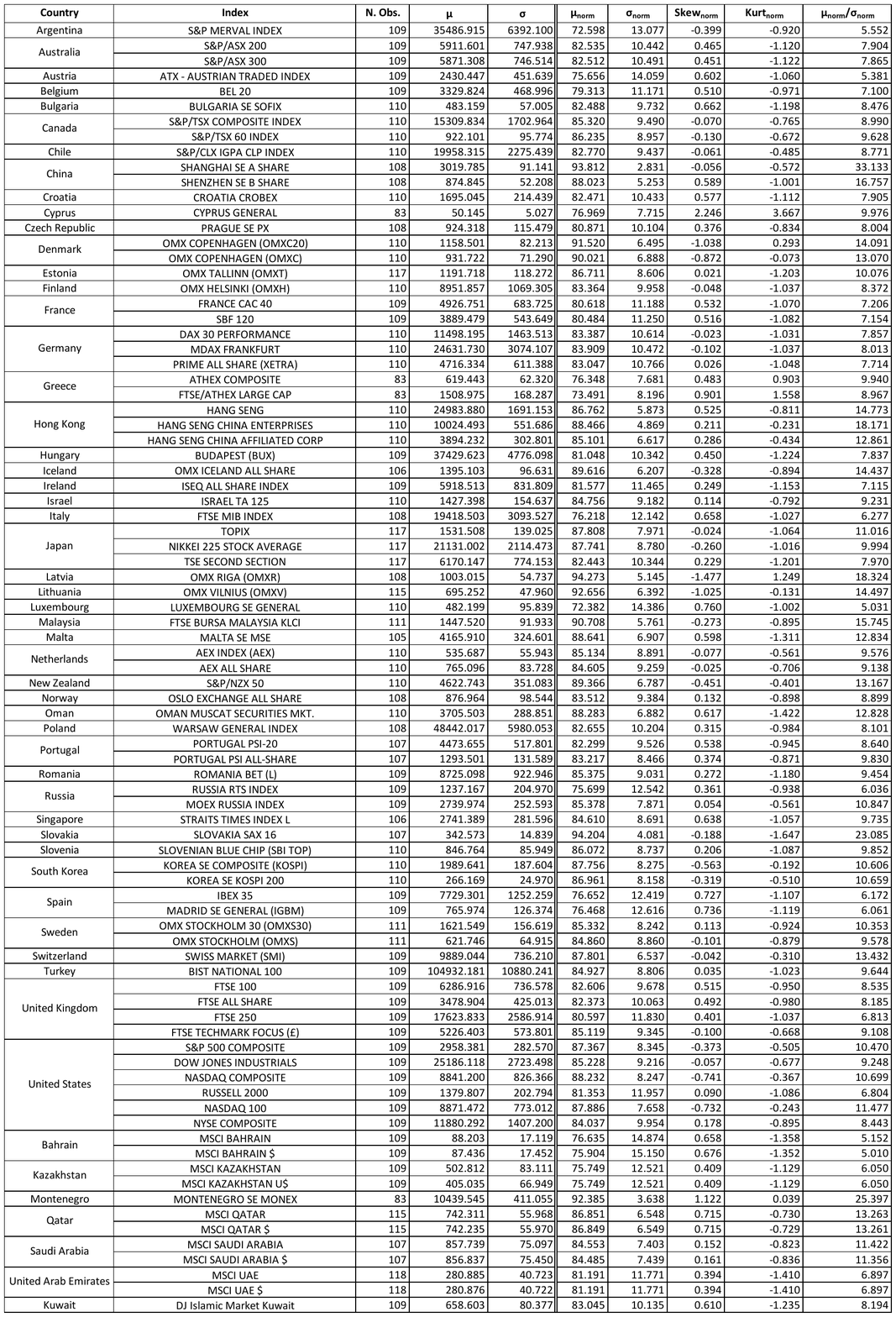}
\caption{The statistical summary of the stock indexes closing prices is reported. The last four columns regard the normalized time series, according the Eq. \eqref{normalizer}.}
\label{tab:2}
\end{table}

\section{Methodology}
\label{method}
To face the problem, we design indicators able to capture the connection between anxiety for the pandemic and expectations on the future outcomes of financial markets. 
%optimism and pessimism generated by COVID-19 addressing Google trends and the financial stock index time series at a country level. 
The underling idea relates to the synchronicity between increments and decrements of Google searches and of stock index levels, so that, increasing (decreasing) volumes of searches and decreasing (increasing) prices are associate to pessimistic (optimistic) moods.

To describe the employed methodology, some notation is needed.

We denote the number of considered countries by $J$ -- and $J$ is 54 for us, see Section \ref{data} -- and label the generic country by $j=1, \dots, J$. Each country hosts $K$ financial markets. The number of financial markets depends on the selected country, so that one should write $K=K(j)$. Such a dependence will be omitted when not necessary. Often, $K>1$ -- i.e. the most part of the considered countries is associated to more than one financial market. However, there are cases of countries with $K=1$. The generic financial market is $k=1, \dots, K$.

As already discussed in Section \ref{data}, we have daily data on prices and Google searches of the word \textit{``coronavirus''} (and its translations) in a common reference period of $T$ days. For country $j$, we denote the available time series of the prices of the stock index $k$ by $\mathbf{p}_k^{j}=(p_k^{j}(1), \dots, p_k^{j}(T))$. Analogously, the sample of the Google searches for country $j$ is $\mathbf{w}^{j}=(w^{j}(1), \dots, w^{j}(T))$.

Notice that the range of variation of the components of $\mathbf{p}_k^{j}$ and $\mathbf{w}^{j}$ are different. Indeed, $\mathbf{p}_k^{j}$ has nonnegative components, while the components of $\mathbf{w}^{j}$ are integer numbers ranging in $[0,100]$, and there exists $\bar{t}$ such that $w^{j}(\bar{t})=100$. Time $\bar{t}$ represents the day with the maximum level of searches over the period $[1,T]$, and depends naturally on $j$. Also such a dependence will be conveniently omitted. The minimum value of the elements of $\mathbf{w}^{j}$ is not necessarily null. Indeed, null search means absence of interest for the considered word in country $j$ -- i.e., null amount of Google searches; such an occurrence does not necessarily appear over the period $[1,T].$ Assigning value 100 to the highest daily flow of Google searches over $[1,T]$ and null value to null searches allows the easy normalization -- implemented directly by the Google Trends algorithm -- of the Google search data in the range $[0,100]$.

For better comparisons, we impose the variation range $[0,100]$ also to the series $\mathbf{p}_k^{j}$ for each $j$ and $k$ through a simple normalization procedure.
We denote the normalized series of the prices by $\bar{\mathbf{p}}_k^{j}$.

First of all, we identify $\bar{t} \in \{1, \dots, T\}$ such that $p_k^{j}(\bar{t})=\max\{p_k^{j}(t):t=1, \dots, T\}$. Then, we set $\bar{p}_k^{j}(\bar{t})=100$. Null price is associated to zero value for the normalized series, so that we set $\bar{p}_k^{j}(t)=0$ when ${p}_k^{j}(t)=0$. Evidently, one can have ${p}_k^{j}(t)>0$ for each $t=1, \dots, T$, so that one has $\bar{p}_k^{j}(t)>0$ for each $t$.

The entire series can be derived as follows
\begin{equation}
\bar{p}_k^{j}(t)=\left[100 \times \frac{{p}_k^{j}(t)}{p_k^{j}(\bar{t})} \right], \qquad \forall\, t=1, \dots, T,
\label{normalizer}
\end{equation}
where $[\bullet]$ is the integer part of the real number $\bullet$.

The exploration and comparison of financial data and Google Trends will proceed at country level; it will be implemented by conceptualizing suitable distance measures, under different perspectives. In so doing, we provide several insights on countries regularities and discrepancies.

\subsection{Time-dependent distance measures}

We first build a distance measures based on the comparison between the time-dependent normalized accumulations of prices and Google searches. We consider $t_1, t_2 \in \{1, \dots, T\}$ with $t_1 \leq t_2$ and define

\begin{equation}
A_j([t_1,t_2];k)=\frac{1}{2}\cdot \sum_{s=t_1}^{t_2} \left[ \frac{\bar{p}_k^{j}(s)}{\bar{P}_k^j}-\frac{w^{j}(s)}{W^{j}} \right]+\frac{1}{2},
\label{A_j}
\end{equation}
where
$$
W^j=\sum_{t=1}^{T}w^{j}(t), \qquad \bar{P}_k^j=\sum_{t=1}^{T}\bar{p}_k^{j}(t).
$$
By construction, it results $A_j([t_1,t_2];k) \in [0,1]$. A high value of $A_j([t_1,t_2];k) $ means that $[t_1, t_2]$ is a time period with a high percentage of price of market $k$ and a low percentage of Google searches -- where percentages have to be intended in terms of the total amount on the overall period. Thus, $A_j([t_1,t_2];k) $ close to one means that $[t_1,t_2]$ is an optimistic period. Differently, $A_j([t_1,t_2];k)$ is close to zero when prices are relatively low and Google searches of the word \textit{``coronavirus''} are relatively high. In this case, $[t_1,t_2]$ is a time interval where country $j$ has experienced anxiety about COVID-19 and lack of trust in market $k$.

Notice that the case $t_1=1$ and $t_2=T$ is trivial and not interesting, being $A_j([1,T];k)=1/2$ for each $j$ and $k$ -- i.e., in the middle (fair) situation between optimism and pessimism. Indeed, $[1,T]$ is the entire period, hence being associated to full percentage of prices and Google searches. More reasonably, the proper selection of $t_1$ and $t_2$ allows to explore elements of the considered sample in relevant subperiods.

At a country level, we can average the $A_j$'s in (\ref{A_j}) with respect to the markets. In particular, we define
\begin{equation}
A_j([t_1,t_2])=\frac{1}{K(j)} \sum_{k=1}^{K(j)}  A_j([t_1,t_2];k).
\label{A_jtot}
\end{equation}
We observe that $A_j([t_1,t_2]) \in [0,1]$, and all the comments reported above remain valid for the indicator presented in (\ref{A_jtot}).

\subsection{Global distance measures}
We here compare the considered series on the basis of the signs of their daily variations. Specifically, we assess how often an increase (a decrease) of the Google searches is associated to a decrease (an increase) of the prices of the financial markets. The entity of the daily variation is also taken into account.

Consistently with our framework, we refer hereafter to a generic series $\mathbf{x}=(x(1), \dots, x(T))$, whose components range in $[0,100]$.

Thus, given a threshold $\zeta \in [0,100]$ and $t=1, \dots, T-1$, we define the sign variation
of the series $\mathbf{x}$ between $t$ and $t+1$ at the threshold $\zeta$ as follows:
\begin{equation}
\label{v} \delta_{t}^{(\zeta)}(\mathbf{x})=\left\{
  \begin{array}{ll}
    1, & \hbox{if $x(t+1)-x(t) > \zeta$;} \\
    0, & \hbox{if  $-\zeta \leq x(t+1)-x(t) \leq \zeta$;} \\
    -1, & \hbox{if  $x(t+1)-x(t) <-\zeta$.} \\
  \end{array}
\right.
\end{equation}

The parameter $\zeta$ is fixed a-priori; it represents the entity of the daily variation to be crossed for stating that the series have an increase (or a decrease, by taking the variation with negative sign) from time $t-1$ to time $t$.
Evidently, the case $\zeta=0$ leads to $\delta_{t}^{(0)}(\mathbf{x})=1$ when $x(t+1)>x(t)$,  $\delta_{t}^{(0)}(\mathbf{x})=-1$ when $x(t+1)<x(t)$ and  $\delta_{t}^{(0)}(\mathbf{x})=0$ when $x(t+1)=x(t)$.

The comparison between the behaviors of the Google searches and of the financial markets can be performed at country level by means of the $\delta$'s defined in (\ref{v}).

For each $j=1, \dots, J$, we compare the series $\mathbf{w}^{j}$ with $\bar{\mathbf{p}}_k^{j}$, for each $k=1, \dots, K(j)$.

We define
\begin{equation}
\label{Delta} \Delta^{(\zeta)}(t,j,k)= \delta_{t}^{(\zeta)}(\mathbf{w}^{j})-\delta_{t}^{(\zeta)}(\bar{\mathbf{p}}_k^{j}).
\end{equation}
By definition, the $\Delta$'s in (\ref{Delta}) can take values in $\{-2,-1,0,1,2\}$. Such values have specific meanings and deserve an interpretation.

When $\Delta^{(\zeta)}(t,j,k)=-2$, then we observe a decrease of the Google searches related to \textit{``coronavirus''} and an increase of the price of the financial market $k$. This case has a clear interpretation in terms of optimism. Indeed, people show a decreasing anxiety for the pandemic disease -- they weaken the amount of searches on the Google -- and, simultaneously, exhibit an increasing interest in investing in the financial market.
The value -1 is associated to constant Google searches and increase of the price or decreasing level of Google searches and invariant price.
The value 0 is related to the cases of identical behavior between Google searches and price, so that they can be invariant between date $t$ and $t+1$ or both of them can increase/decrease.
The value +1 relies to increasing level of Google searches and invariant price or, alternatively, a constant level of Google searches and decreasing price.
The value +2 describes the situation in which Google searches grow and price decrease. This is the other corner case, in which anxiety and sadness for the spread of the disease -- mirroring in the growth of Google searches -- is associated to decreasing amount of investments in the financial market.

In general, the positive values of the $\Delta$'s describe situations of pessimism, captured by anxiety for the disease and decrease of investments in the financial markets. Conversely, the cases of negative $\Delta$'s are related to optimism, with decreasing interest for COVID-19 and growing attention for the future evolutions of financial markets.

Some distance measures with high information content can be derived by (\ref{Delta}).

We measure the aggregated connection between the considered trend in Google and the price of market $k$ in country $j$ over the considered period by defining
\begin{equation}
H_j^{(\zeta)}(k)=\frac{1}{4(T-1)}\left[\sum_{t=1}^{T-1} \Delta^{(\zeta)}(t,j,k)+2(T-1)\right].
\label{H_j}
\end{equation}
By construction, $H_j^{(\zeta)}(k) \in [0,1]$. If such an indicator approaches zero, then people in country $j$ tend to the highest level of optimism -- in the sense expressed when discussed the case of -2 as value of the $\Delta$'s  -- when analyzing the Google searches of the considered word and its connections with the price of financial market $k$. The converse situation appears when $H_j^{(\zeta)}(k)$ is close to one, where we are in presence of a high level of pessimism and anxiety.

By averaging the $H_j$'s in (\ref{H_j}) with respect to $k$ we obtain an indicator describing the reality at country level, for all the connections between the considered word and the prices of financial markets, as follows:
\begin{equation}
\label{H_jtot} H_j^{(\zeta)}=\frac{1}{K(j)}\sum_{k=1}^{K(j)}  H_j^{(\zeta)}(k).
\end{equation}
Clearly, $ H_j^{(\zeta)} \in [0,1]$ and the arguments above -- opportunely rephrased at country level -- remain valid.

We now provide a measure of how a specific country has experienced optimism versus pessimism over the considered period. At this aim, we consider a ratio indicator as follows:
%\begin{equation}
%R_j^{(\zeta)}(k)=\frac{1}{T-1/T}\cdot \left[\frac{1+\sum_{t=1}^{T-1} \mathbf{1}\left( \Delta^{(\zeta)}(t,j,k)=-2\right)}{1+\sum_{t=1}^{T-1}  \mathbf{1}\left( \Delta^{(\zeta)}(t,j,k)=2\right)}-\frac{1}{T} \right],%\sum \Delta^{(\zeta,\mathbf{p})}(:,j,h,k),
%\label{R_j}
%\end{equation}

\begin{equation}
R_j^{(\zeta)}(k)=\frac{1}{2(T-1)}\left[ \sum_{t=1}^{T-1} \mathbf{1}\left( \Delta^{(\zeta)}(t,j,k)=2\right)-\sum_{t=1}^{T-1}  \mathbf{1}\left( \Delta^{(\zeta)}(t,j,k)=-2\right) +T-1 \right].
\label{R_j}
\end{equation}
where
$$
 \mathbf{1}(\bullet)=\left\{
  \begin{array}{ll}
    1, & \hbox{if $\bullet$ is true;} \\
    0, & \hbox{otherwise.} \\
  \end{array}
\right.
$$
By construction, $ R_j^{(\zeta)}(k) \in [0,1]$. For country $j$ and when referring to market $k$, there is a high percentage of optimistic days with respect to pessimistic ones as the value of such indicator approaches zero, while we are in a substantial context of pessimism when the indicator in (\ref{R_j}) is close to one. The corner cases have a clear interpretation: when $ R_j^{(\zeta)}(k) =0$, then all the days in the considered period present a decreasing anxiety for COVID-19 coupled with an increasing trust in market $k$; differently, $ R_j^{(\zeta)}(k) =1$ is associated to an entire period of increasing need of awareness on COVID-19 and decreasing price of market $k$.

Also in this case, we can focus on country $j$ by averaging the $R_j$'s over the markets:
\begin{equation}
\label{R_jtot} R_j^{(\zeta)}=\frac{1}{K(j)}\sum_{k=1}^{K(j)}  R_j^{(\zeta)}(k).
\end{equation}
Evidently, $R_j^{(\zeta)} \in [0,1]$ and the discussion reported above applies also in this more general case.

The global distance measures presented above capture two different aspects of the phenomenon under analysis. $H_j^{(\zeta)}$ and $H_j^{(\zeta)}(k)$ provide information on the mood as an average of the $\Delta$'s over all the days of the considered sample. Differently, $R_j^{(\zeta)}(k)$ and $R_j^{(\zeta)}$  focus only on the dates where the daily variations of searches volumes and stock index levels have had discordant behaviors. Namely, the indicators $R$'s offer more details on the ratio between fully optimistic days and fully pessimistic ones, i.e. on the proportion of the days in which the Google searches have decreased and the indexes prices have increased and those with an increase of searches and a decrease of the prices.

\begin{figure}[!htb]
\includegraphics[width=\textwidth]{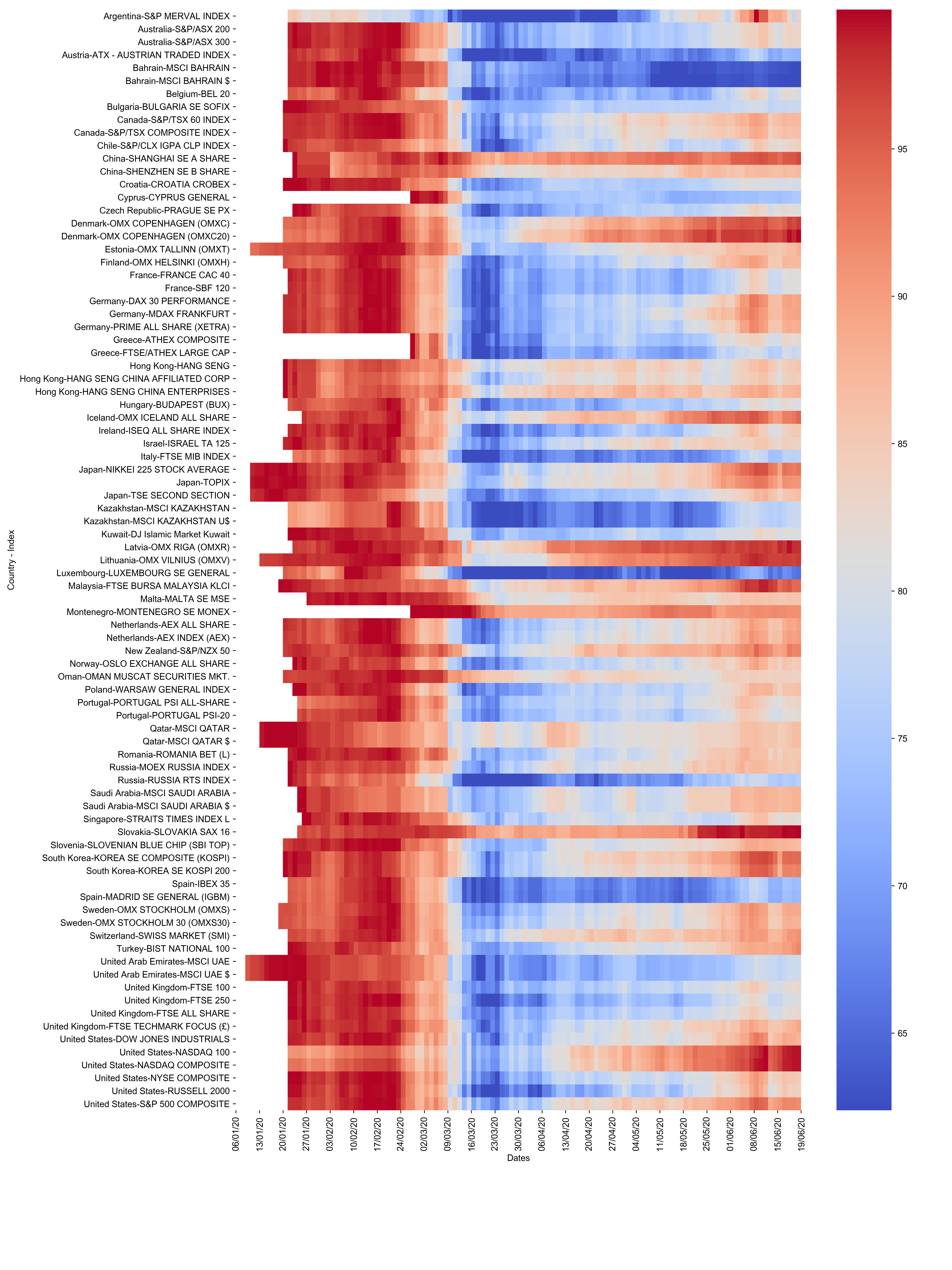}
\caption{Heatmap representation of the normalised prices recorded for each stock market index (see, Eq. \ref{normalizer}). The time series starting points are different because the prices are stored from the first day in which relevant volumes of Google searches in that country are recorded.}
\label{fig:2}
\end{figure}

\section{Results and discussion}
\label{results}
The normalised time series of the stock indexes prices are obtained via Eq. \eqref{normalizer}. The outcome of such a normalisation is presented in Figure \ref{fig:2} and the main statistical indicators of both the original and the normalised time series are showed in Table \ref{tab:2}. The visual inspection of this Figure allows the reader to confirm the general trends of the markets, with a decline inducted by incorporation of the pandemic effects. Figure \ref{fig:1} and Table \ref{tab:1} show the increased Google searches of the translated \textit{``coronavirus''} in different countries. The searching activities started at a different time and with a general delay with respect to the decline recorded in the financial markets.

As a preliminary comment, we notice that the Moreover, the $A_j$'s in Eq. \eqref{A_j} and \eqref{A_jtot} compare the normalised values of Google searches and prices, while the $H_j$'s in Eq. \eqref{H_j} and \eqref{H_jtot} and the $R_j$'s in Eq. \eqref{R_j} and \eqref{R_jtot} compare their daily increments and decrements. Thus, the $A_j$'s offer a view of the snapshots of anxiety for COVID-19 and trust in financial markets; differently, the $H_j$'s and the $R_j$'s propose an evolutive perspective on the daily variations of the Google search and the stock market data.

\begin{figure}[!htb]
\includegraphics[width=\textwidth]{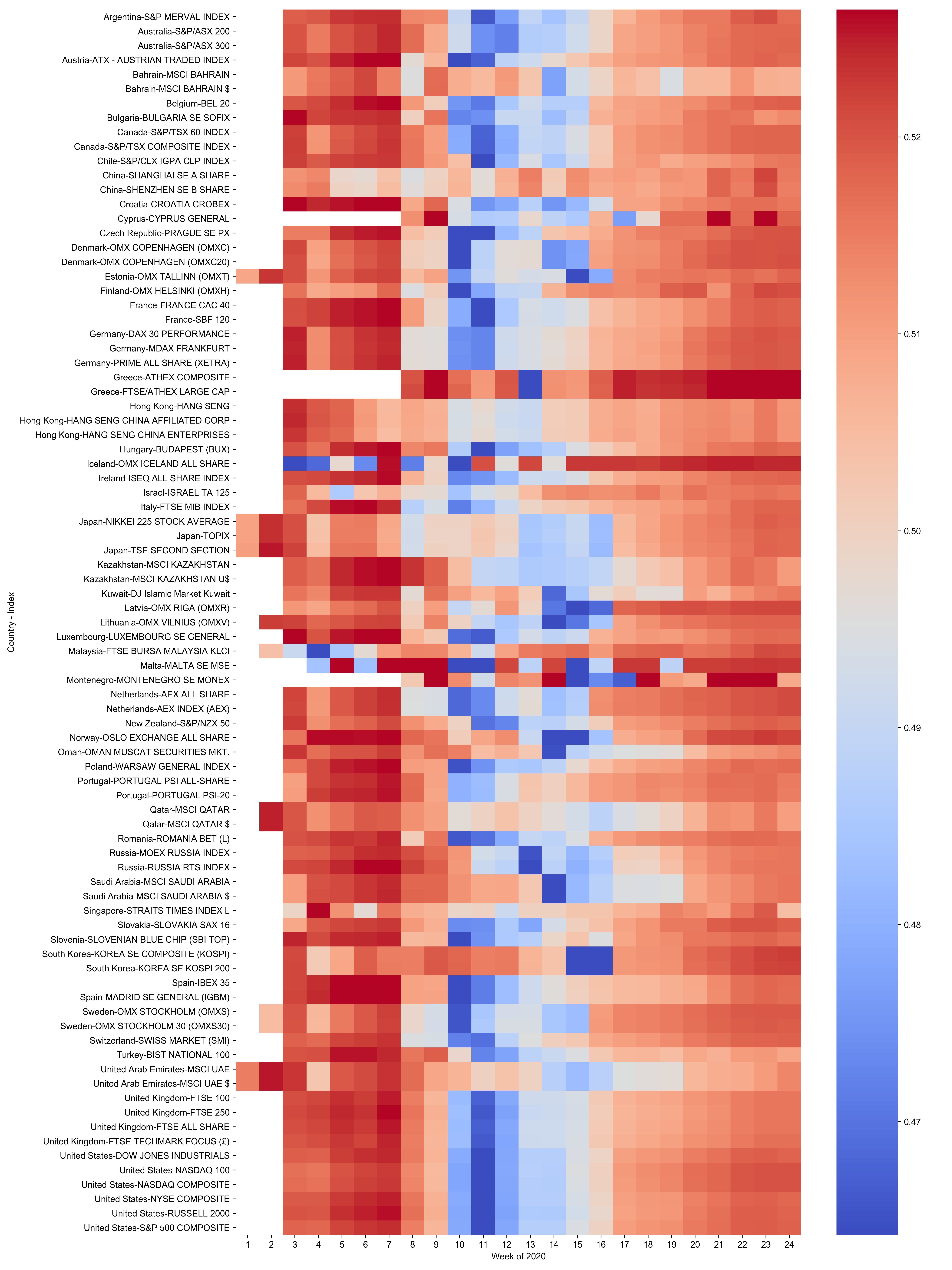}
\caption{Heatmap representation of $A_j([t_1,t_2];k)$ in Eq. \eqref{A_j}, at stock index level. Indents represent the differences in the starting date of the related Google Trends data -- i.e., the first date with a nonnull Google search volume.}
\label{fig:3}
\end{figure}

\begin{figure}[!htb]
\includegraphics[width=\textwidth]{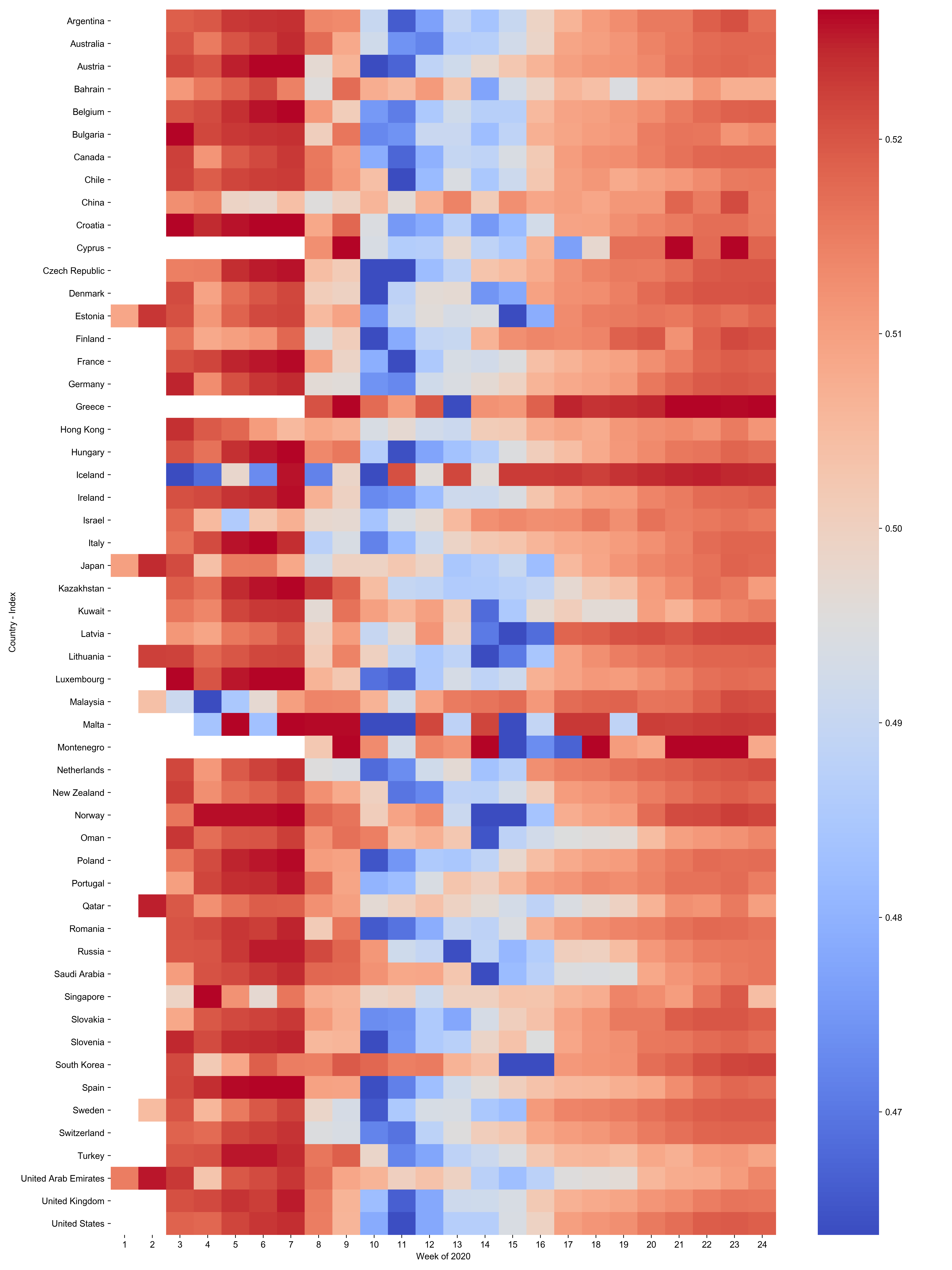}
\caption{Heatmap representation of $A_j([t_1,t_2])$ in Eq. \eqref{A_jtot}, at country level. Also in this case, indents represent the differences in the starting date of the Google Trends data at country level.}
\label{fig:4}
\end{figure}

In computing the index $A_j([t_1,t_2];k)$ in Eq. \eqref{A_j}, we take $t_2 - t_1$ constantly equal to five days, hence studying the weekly behaviour of the index. The outcomes per each index are summarized in Figure \ref{fig:3} and Table \ref{tab:3}. Moreover, the results of $A_j([t_1,t_2])$ across the stock indexes of each country -- namely, those in Eq. \eqref{A_jtot} -- are reported in Figure \ref{fig:4} and Table \ref{tab:4}. From this view, some facts emerge:
\begin{itemize}
\item The paths have drastically changed between the $7^{th}$ and the $8^{th}$ weeks of the year, namely between 17/02/2020 and 01/03/2020. This is the period during which the international community started to take the situation seriously despite the controversial statements of national governments' heads. On 11/03/2020, WHO's Director declared ``WHO has been assessing this outbreak around the clock and we are deeply concerned both by the alarming levels of spread and severity, and by the alarming levels of inaction. We have therefore made the assessment that COVID-19 can be characterised as a pandemic.'' \cite{world2020director}.

\item Greece and South Korea have spent more than $90\%$ of the analysed weeks in a quite positive mood, namely reporting an $A_j([t_1,t_2]) > 0.5$.

\item Cyprus and Iceland have experienced mild pessimism on a quite large number of weeks. They present $A_j([t_1,t_2]) < 0.5$ at least $40\%$ of the times in the studied period.

\item Weeks 10 and 11 are characterized by the lowest average of $A_j([t_1,t_2])$. Their means across the countries are respectively 0.485 and 0.483.

\item The highest number of countries experiencing a $A_j([t_1,t_2]) < 0.5$ is met on week 11. During the period 16/03/2020 - 20/03/2020, $81\%$ of the analysed countries experienced a high volume of Google searches and a low level of normalised prices. Therefore, a high level of anxiety/pessimism. On the other hands, the tails (weeks 1-4 and 20-24) present a higher level of the index, with an increased presence of positivism in most of the countries during the most recent weeks.

\end{itemize}
In Table \ref{tab:9} the considered countries are week-wise ranked by using $A_j([t_1,t_2])$. Montenegro holds the first position for five weeks. Similarly, we observe that Greece, Iceland and Malta seat on the firsts four positions most of the times.  This outcome suggests that these countries experienced waves of optimism and pessimism; interestingly, for the quoted countries, consecutive weeks may have a large discrepancy in the ranking positions. Thus, one can say that the waves are of impulsive and compulsive nature -- perhaps, they are driven by news on the pandemic or statements of the Governments -- and this leads to sudden changing of the people's behaviour in searching on Google and taking positions in the market.

\begin{figure}[!htb]
\includegraphics[width=\textwidth]{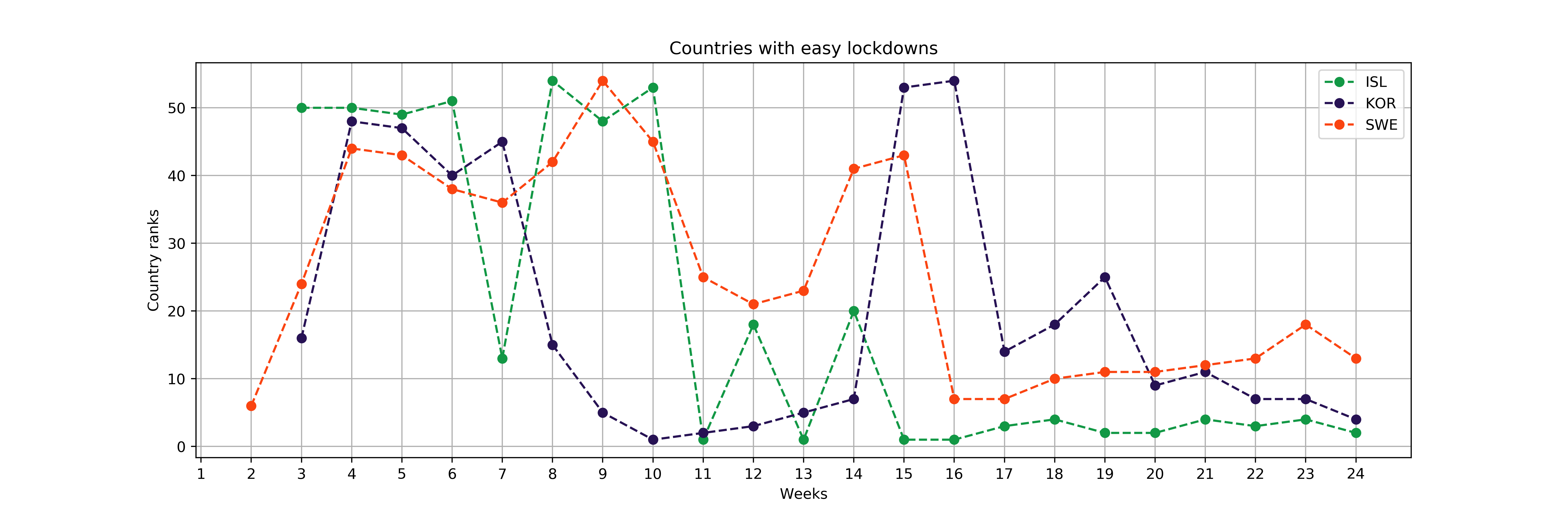}
\includegraphics[width=\textwidth]{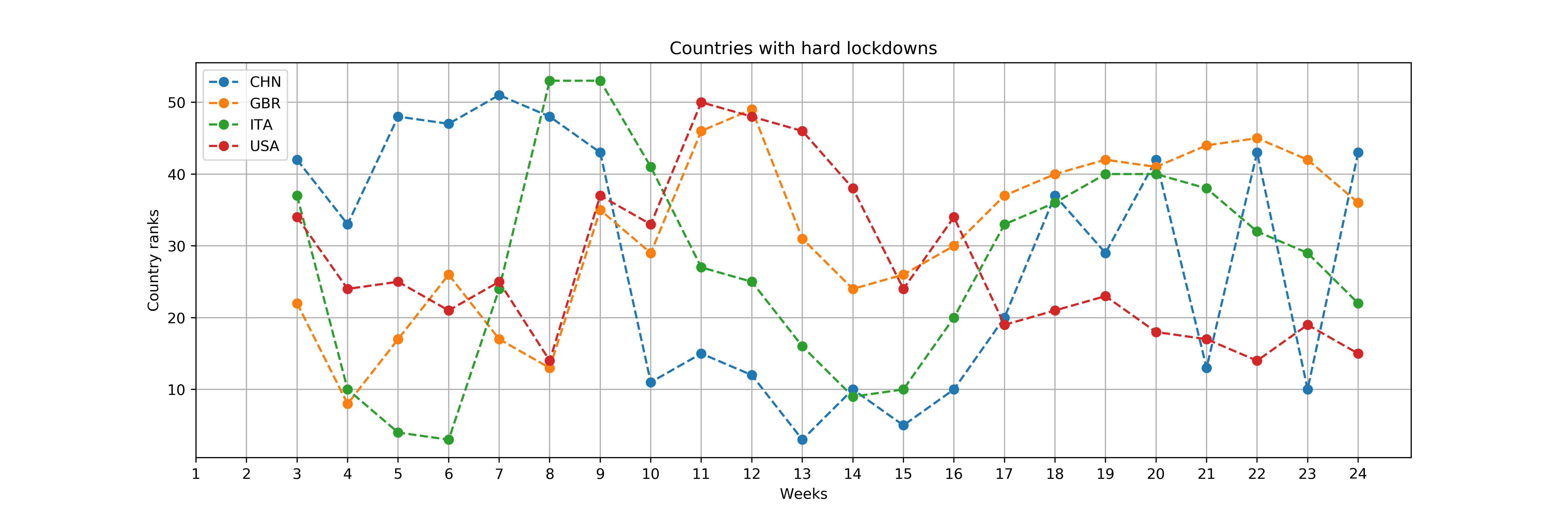}
\caption{A comparison of the weekly mood of the countries -- on the basis of the ranks of $A_j([t_1,t_2])$ -- having experienced an easy/hard lockdown. The lower is the rank, the higher is the optimism experienced in that week by the respective country.}
\label{fig:10}
\end{figure}

We also propose a focus of weekly rankings of some paradigmatic cases: Sweden, Iceland and South Korea -- the countries with an easy lockdown, see \cite{wiki:ISL,wiki:SWE,KOR} -- and Italy, UK, USA and China -- which are countries having or having had a harder lockdown. By inspecting Figure \ref{fig:10}, one can appreciate that the countries having experienced an easier lockdown have spent more optimistic moods during the recent weeks.

\begin{figure}[!htb]
\includegraphics[width=\textwidth]{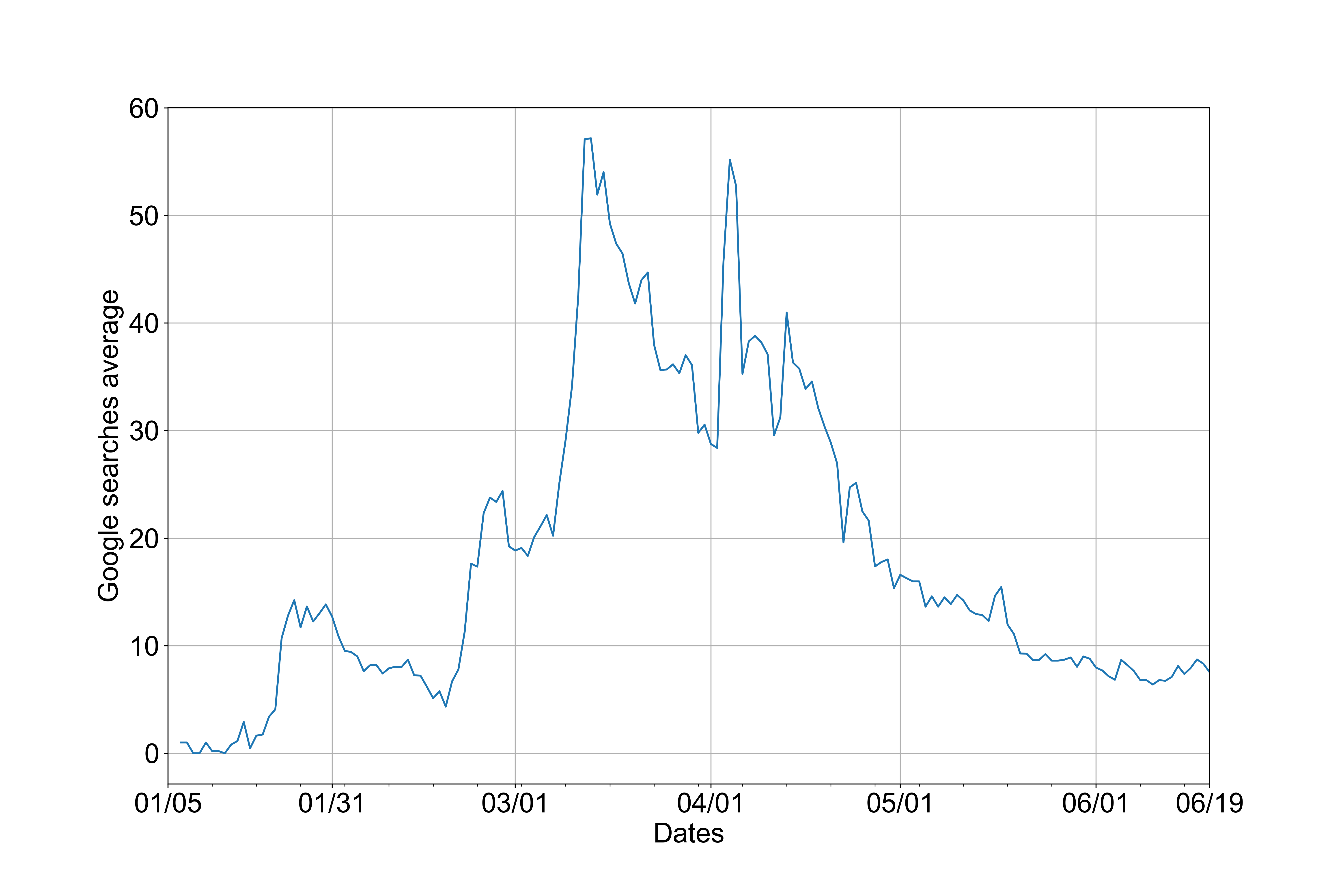}
\caption{Averaged Google searches of \textit{``coronavirus''} -- along with its translations in the different languages -- across countries with $HDI > 0.8$ plus China, on time-basis.}
\label{fig:9}
\end{figure}

The results show some regularities in the behavior across countries and indexes, as Figures \ref{fig:3} and \ref{fig:4} clearly testify. An initial phase of optimism was probably induced by skeptic statements from national governments and media agencies; in fact, the emergence has been underestimated by a large number of people at its inception, see \cite{colarossi2020}. Then, once the situation has escalated, Google searches have drastically increased (see Figure \ref{fig:9}) and the markets have simultaneously reacted, plausibly also in the light of the lockdown policies implemented all over the world. The raised pessimism is represented in Figure \ref{fig:3} and \ref{fig:4} by the blue bands in weeks 10-15. A general relief came in after that. In a few cases, the anxiety boosted from the very beginning. This is clearly the case for Iceland, Malaysia, Malta and more mildly for Singapore, see Figure \ref{fig:4} and Table \ref{tab:9}.
Considering week $24^{th}$, the stock indexes and so the countries reporting the highest level of $A_j$ in \eqref{A_jtot} are Greece, Iceland and Malta, with values 0.527, 0.524, 0.523, respectively. On the other hand, those having the lowest values are Montenegro, Bahrain and Singapore with 0.508, 0.507 and 0.504, respectively. %Interestingly, all the values are above 0.5. Therefore, all the countries are into a slightly optimistic mood. 

Figure \ref{fig:10} offers a comparison of the weekly rank of the countries -- on the basis of $A_j([t_1, t_2])$ -- having experienced an easy (upper panel) and hard (lower panel) lockdown. 
In general, countries with a stricter lockdown seem to show globally a more pervasive pessimistic moods than those with a weaker lockdown.
In particular, one can notice the presence of common waves of optimism (low rank) and pessimism (high rank) over the considered period. Importantly, there is an evident countertendency among some countries, with opposite moods in peculiar subperiods. Indeed, Iceland, South Korea and Sweden show pessimism at the beginning of the pandemic and optimism for the rest of the period, with a spike of pessimism around week 15-16. For China, UK, Italy and USA the situation is more scattered, but there is optimism at the beginning for UK, Italy and USA, a substantial pessimism of all the considered countries in the last part of the period.  China and Italy seem to follow analogous patterns in the late part of the period; a possible explanation can be found in the strict collaboration between such countries during the lockdown, which can be seen as the driver of a common mood.

Eqs. \eqref{H_j} and \eqref{R_j} allows getting the global distance measures considering different levels of $\zeta$, which is the threshold used to capture the variations of the observed series on a daily basis. Specifically, we use $\zeta=0,1,\dots, 50$.

\begin{figure}[!htb]
\includegraphics[width=\textwidth]{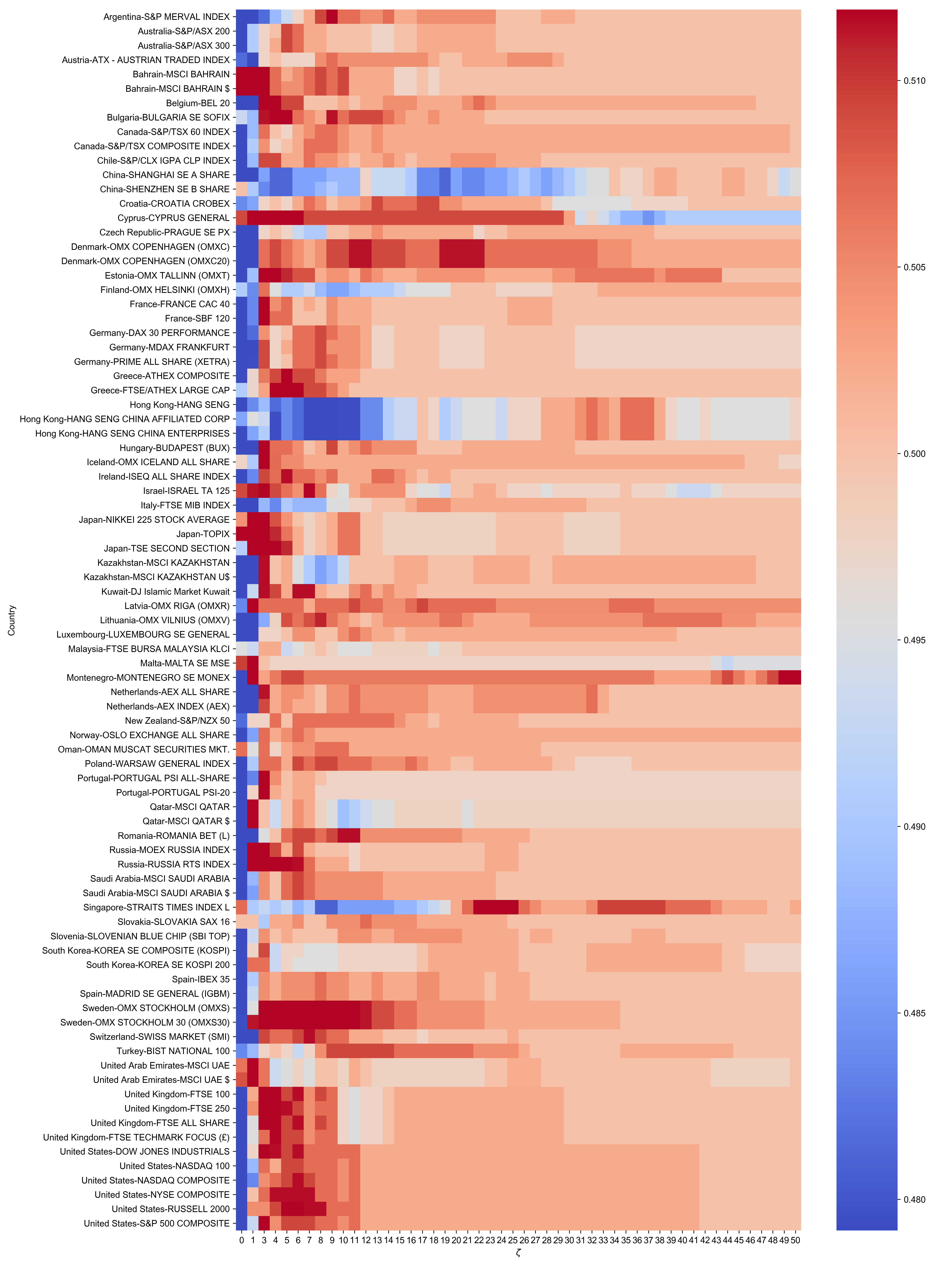}
\caption{Heatmap representation of $H_j^{(\zeta)}(k)$ in Eq. \eqref{H_j}, at stock index level and on the basis of the thresholds $\zeta$s.}
\label{fig:5}
\end{figure}

The results for $H_j^{(\zeta)}(k)$ (see Eq. \ref{H_j}) are reported in Figure \ref{fig:5} and Table \ref{tab:5}.

Financial markets show quite similar behaviours in their links with the Google Trends, mainly in the maximum values of $H_j^{(\zeta)}(k)$. Indeed, the variation range in the maxima is 0.502-0.530, with Bahrain's stock indexes being outliers with 0.551 and 0.567. However, there are noticeable differences in the minimum values of the $H_j^{(\zeta)}(k)$, with a variation range 0.400-0.498. Notice the differences appearing also within the same country, like for the minima of the $H_j^{(\zeta)}(k)$ for the US -- with NYSE COMPOSITE at 0.468 and NASDAQ 100 and NASDAQ COMPOSITE at 0.403.

\begin{figure}[!htb]
\includegraphics[width=\textwidth]{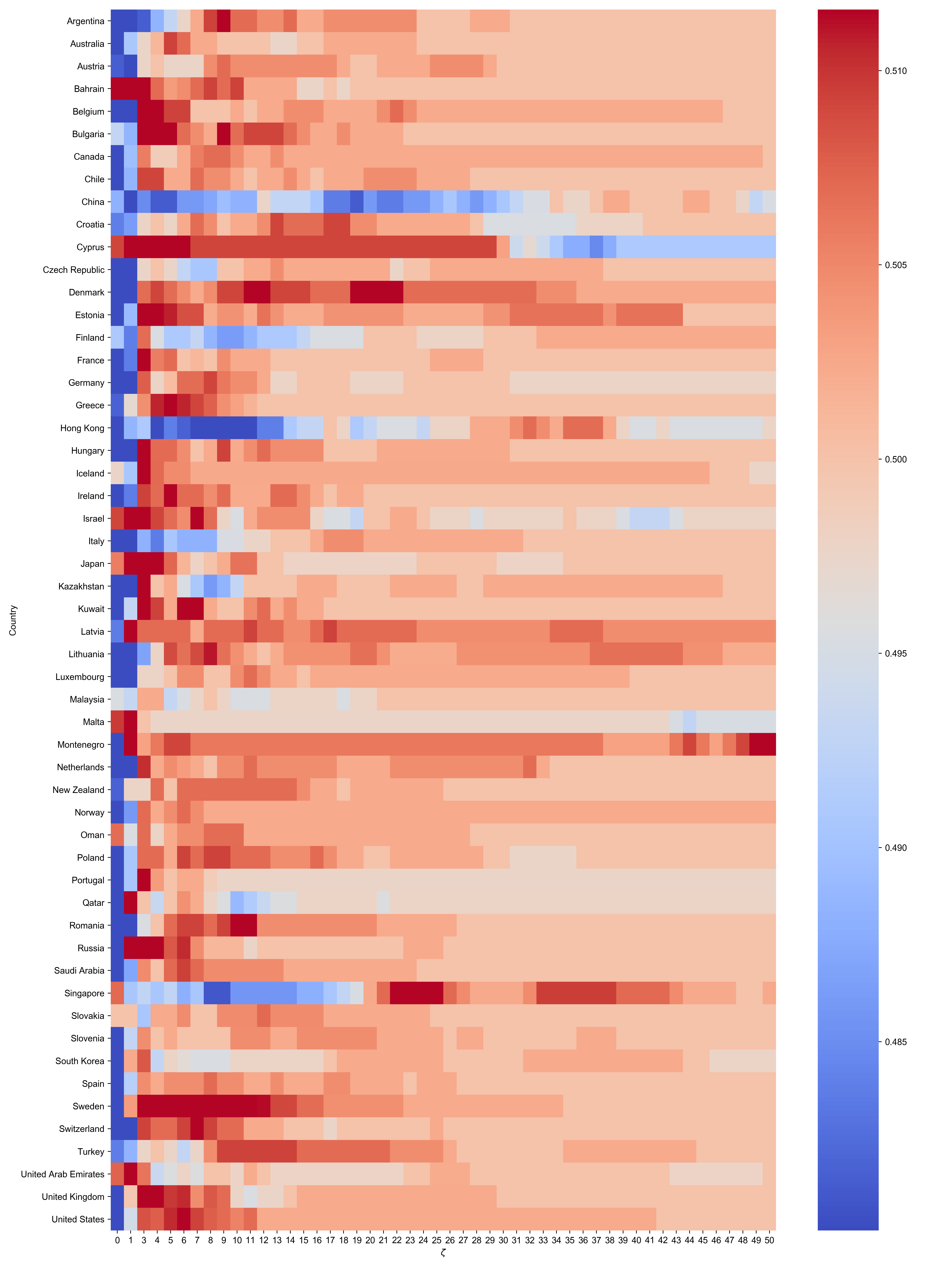}
\caption{Heatmap representation of $H_j^{(\zeta)}$ in Eq. \eqref{H_jtot}, at country level and on the basis of the thresholds $\zeta$s.}
\label{fig:6}
\end{figure}

The averaged results at country level of Eq. \eqref{H_jtot} are shown in Figure \ref{fig:6} and Table \ref{tab:6}.

\begin{figure}[!htb]
\includegraphics[width=\textwidth]{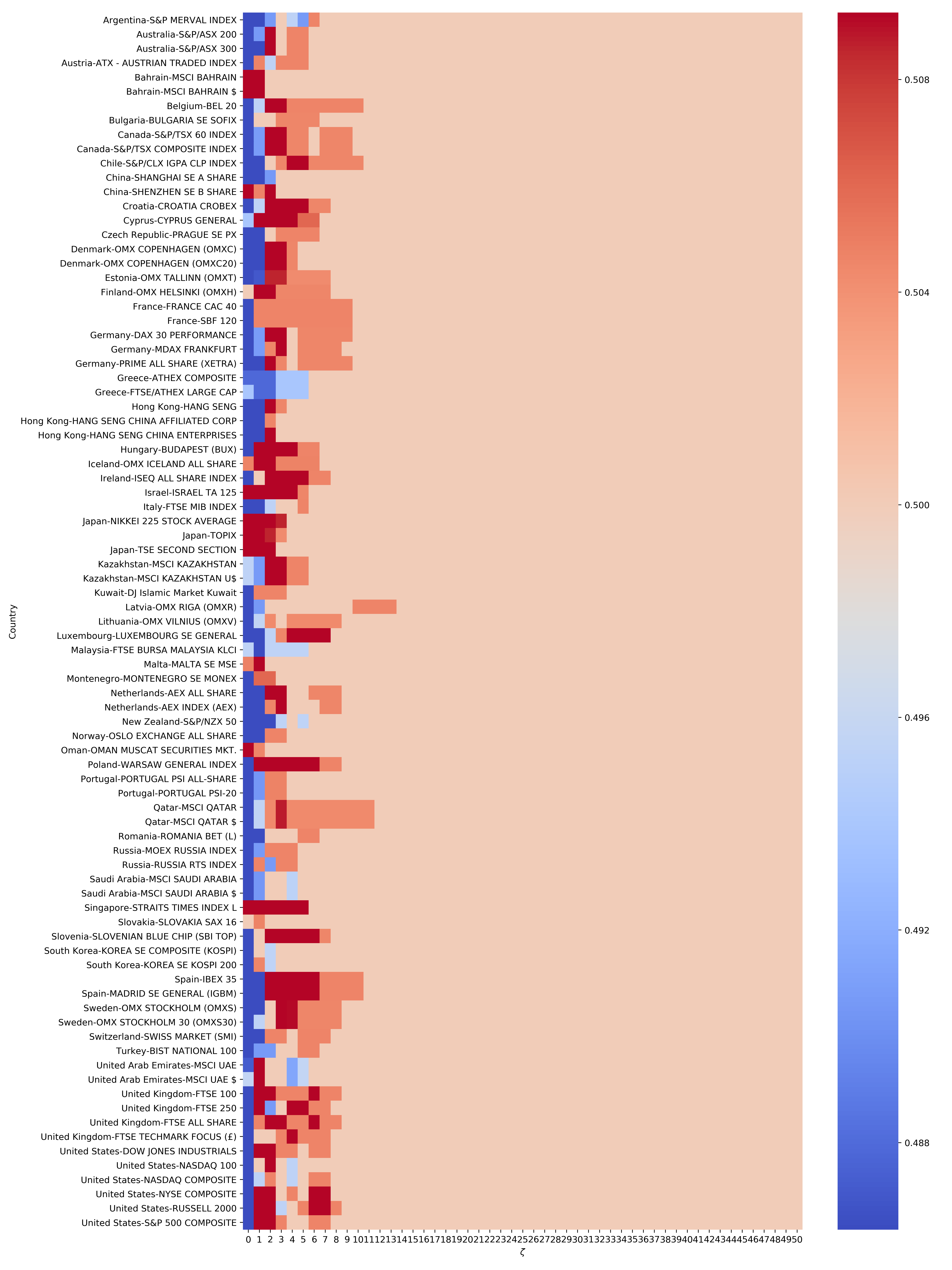}
\caption{Heatmap representation of $R_j^{(\zeta)}(k)$ in Eq. \eqref{R_j}, at stock index level and on the basis of the thresholds $\zeta$s.}
\label{fig:7}
\end{figure}

%Figures \ref{fig:5} and \ref{fig:2}, and Table \ref{tab:5} testify that within countries, when multiple stock indexes are reported, the paths are quite similar if not identical, leaving to the Google Trends searches the leadership in characterizing $H_j^{(\zeta)}$ at country level.  The index is mainly characterized by the distribution of the Google Trends and normalized stock indexes prices daily differences. Their distributional features affect the $H_j^{(\zeta)}$ when $\zeta$ varies.
Some cases are particularly interesting and can be noticed by visual inspecting the results:
\begin{itemize}
\item{Latvia, Montenegro, Norway, Denmark and Canada have a vast majority of $H_j^{(\zeta)} > 0.5$ manifesting a high average level of simultaneous Google searches growths and stock indexes declines. Across the $\zeta$s used in calculating $H_j^{(\zeta)}(k)$, such an occurrence appears at least in the $90\%$ of the cases.}
\item{Malta have $92\%$ of $H_j^{(\zeta)} < 0.5$, representing an average low level of decreasing Google searches and stock indexes increments at the same time.}
\item{The highest value of $H_j^{(\zeta)}$ occurs in Bahrain, with 0.559, for $\zeta=0$. This finding is in agreement with those discussed already for $H_j^{(\zeta)}(k)$ above.}
\item{The smallest value of $H_j^{(\zeta)}$ occurs in Italy, with 0.400, for $\zeta=0$.}
\end{itemize}
The $R_j^{(\zeta)}(k)$ in Eq. \eqref{R_j} are reported in Figure \ref{fig:7} and Table \ref{tab:7}.

The variation range in the maxima for the case of $R_j^{(\zeta)}(k)$ is 0.5 -- 0.565, with Bahrain's stock indexes having the highest values. Differences in the minimum values are noticeable as well, and the variation range goes from 0.421 to 0.5. The lowest value is associated with Italy's index once again. Remarkable differences appear for the markets within the same country, in the specific case of $R_j^{(\zeta)}(k)$; the USA is again one of the most remarkable examples of a wide variation range at a stock market level.

\begin{figure}[!htb]
\includegraphics[width=\textwidth]{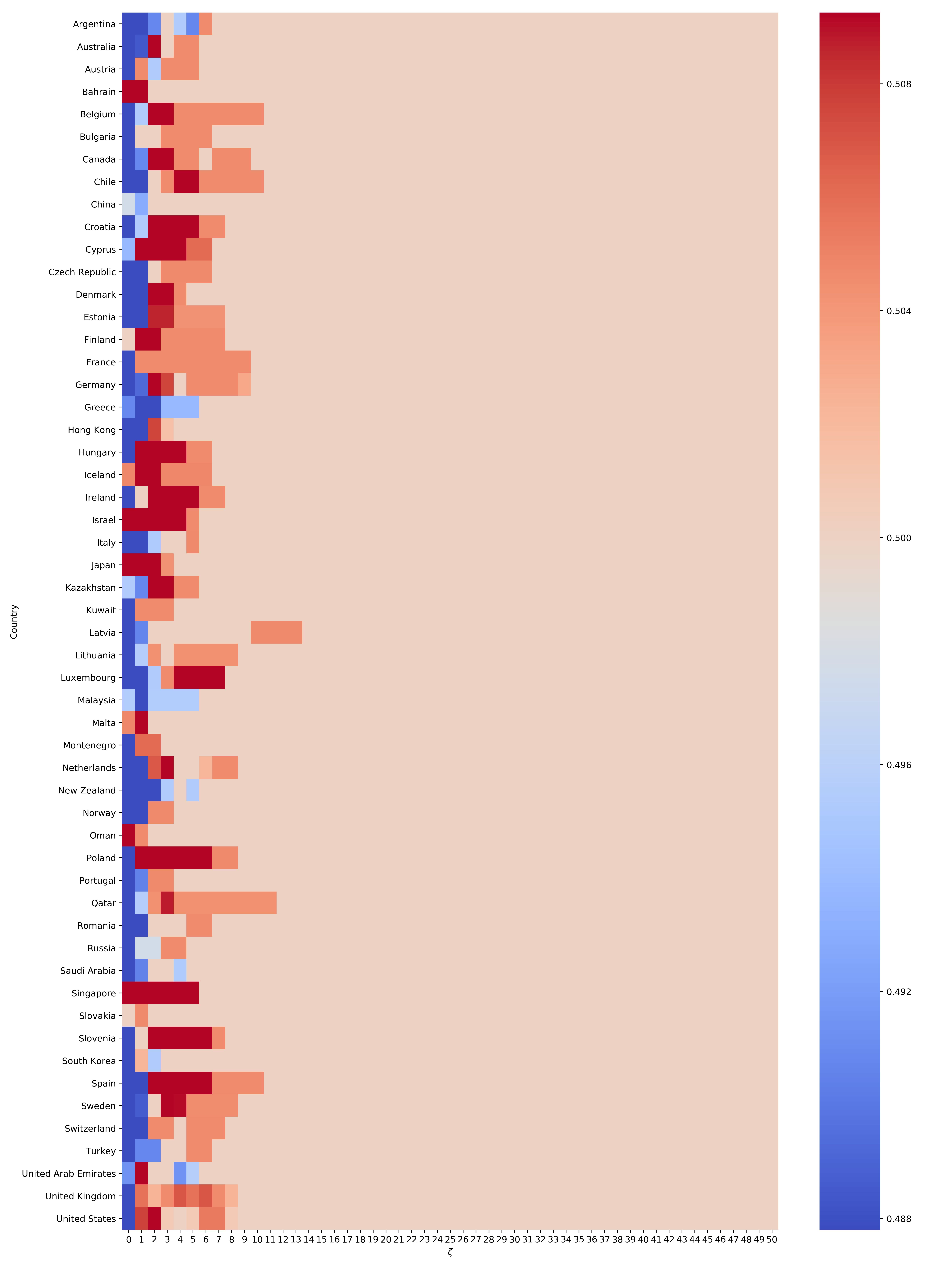}
\caption{Heatmap representation of $R_j^{(\zeta)}$ in Eq. \eqref{R_jtot}, at country level and on the basis of the thresholds $\zeta$s.}
\label{fig:8}
\end{figure}

The results at country level are presented in Figure \ref{fig:8} and Table \ref{tab:8}; they have been calculated through Eq. \eqref{R_jtot}. The most relevant facts are listed below:
\begin{itemize}
\item Qatar has the highest percentage of $\zeta$s such that $R_j^{(\zeta)} > 0.5$, namely $19.6\%$; therefore, it is the country having contemporaneous increases in Google searches and decreases in stock index prices for a large number of thresholds $\zeta$s. Belgium, Spain and France follow, with $17.6\%$ of the $\zeta$s leading to $R_j^{(\zeta)}$ in the range (0.5,1]. %Figure \ref{fig:8} provide the reader with a visual idea.

\item Greece, Malaysia, Argentina and New Zealand have the highest percentages of $\zeta$s such that  $R_j^{(\zeta)} < 0.5$, with the first two countries having $11.8\%$ of the observations falling within [0,0.5) and the latest two ones having a proportion of $9.8\%$. %They manifest an increasing level of Google searches and decreasing prices in their indexes. 

\item The lowest value of $R_j^{(\zeta)}$ occurs in Italy, with 0.421, for $\zeta = 0$. 

\item The highest value of $R_j^{(\zeta)}$ occurs in Bahrain, with 0.565, for $\zeta = 0$.
\end{itemize}

By looking at the global distances measures, the case of $\zeta = 0$ is the most relevant to be commented for the information carried out. In such a case, the proposed indexes are sensible to the smallest daily variation. Bahrain, Malta, Israel, Cyprus, United Arab Emirates, Singapore, Oman and Japan have $H_j^{(\zeta=0)}>0.5$. Thus, these countries have experienced on average a great level of anxiety for COVID-19 and a small trust in the financial markets future performances. Differently, Italy, Canada, Lithuania, Germany, United Kingdom and Spain have the lowest positions, with $H_j^{(\zeta=0)}<0.5$. In such countries, an optimistic mood seems to be preponderant, on average. Notice that such a list of ``optimistic moods'' contains highly developed countries with a noticeable spread of the pandemic. Reasonably, optimism is linked to the confidence of the citizens of the most developed countries either in finance as well as in the health care infrastructures, in the light of solving a so pervasive problem like a widespread pandemic.  %For example, in countries like Italy the virus has spread more rapidly, the governments were not timely responsive and the health care infrastructure was not that capacious and needed time to catch-up the required standards to face the emergency. They have experienced rising Google searches volumes because of the anxiety felt by the citizens. In addition to that, the markets have incorporated the effect of the tough lockdowns experienced by most of these countries. Despite these factors, their mood indicators result slightly positive. A plausible explanation for that can be found in the boost had once the lockdowns have been eased.

For the case of $R_j^{(\zeta=0)}<0.5$, the lowest positions are held by Russia, Switzerland, Lithuania, Romania, Germany and Italy. So, these countries that have experienced a large number of days of contemporaneous decreases in Google searches and increases in stock index prices. Bahrain, Israel, Japan, Singapore, Oman, Malta and Iceland are the countries with $R_j^{(\zeta=0)}>0.5$. Of course, results for $H_j^{(\zeta=0)}$ and $R_j^{(\zeta=0)}$ are often overlapping, and some countries confirm their general mood when the comparison between fully optimistic days and fully pessimistic ones is performed. Interestingly, we find that in places where the pandemic has being managed quite brightly, the general feelings have been more pessimistic than optimistic (see e.g. the case of Israel).

\begin{table}[!htb]
\centering
\includegraphics[scale = 0.75]{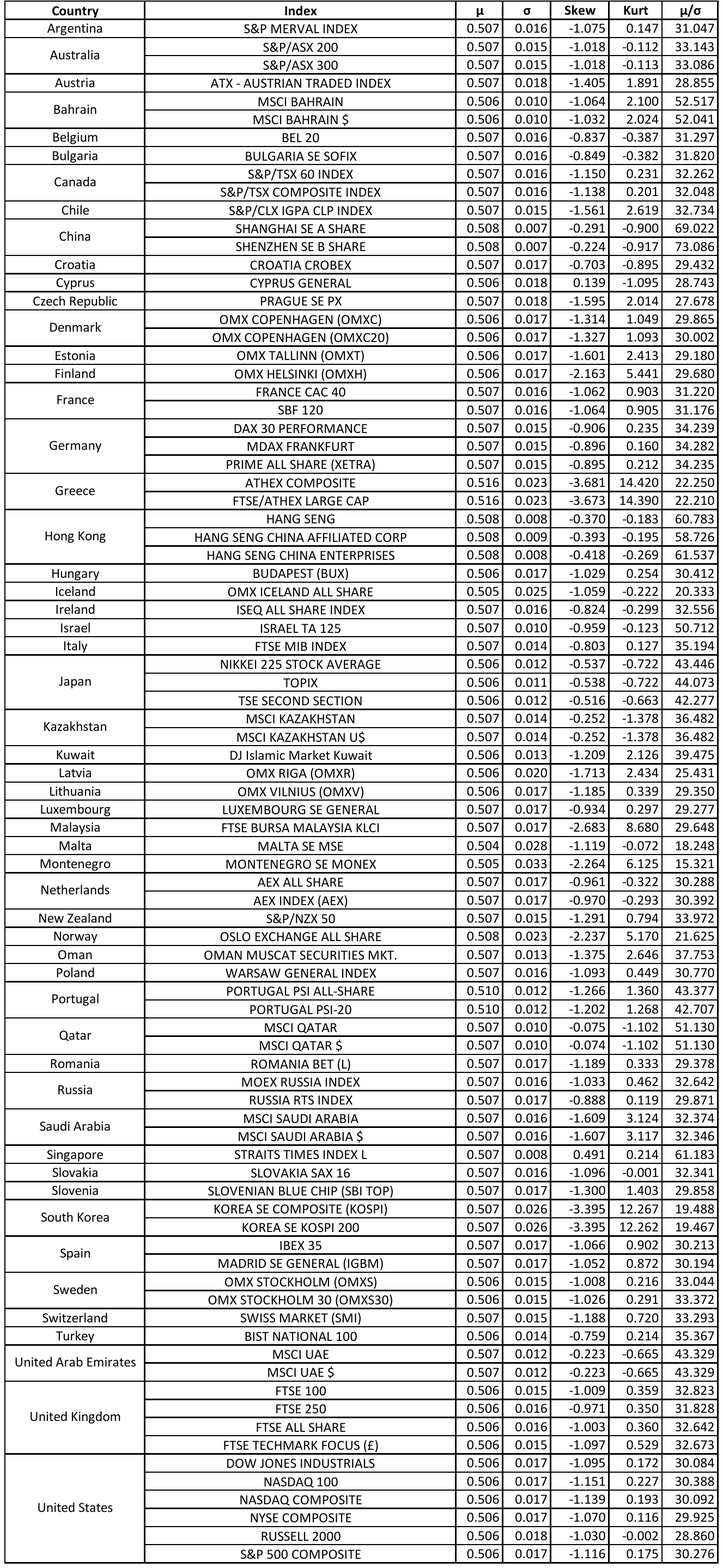}
\caption{Main statistical indicators of $A_j([t_1,t_2];k)$ from Eq. \eqref{A_j} at stock index level.}
\label{tab:3}
\end{table}

\begin{table}[!htb]
\centering
\includegraphics[scale = 0.75]{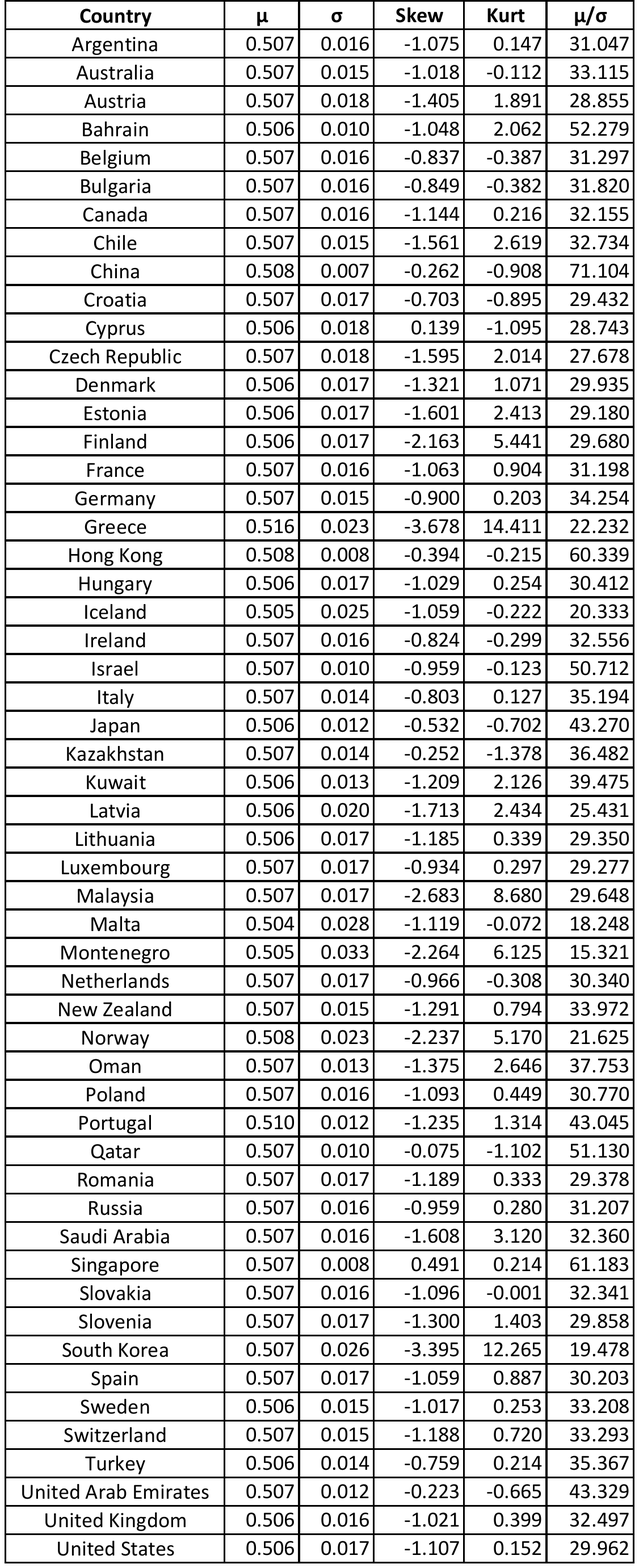}
\caption{Main statistical indicators of $A_j([t_1,t_2])$ in Eq. \eqref{A_jtot} at country level.}
\label{tab:4}
\end{table}

\begin{table}[!htb]
\centering
\includegraphics[scale = 0.75]{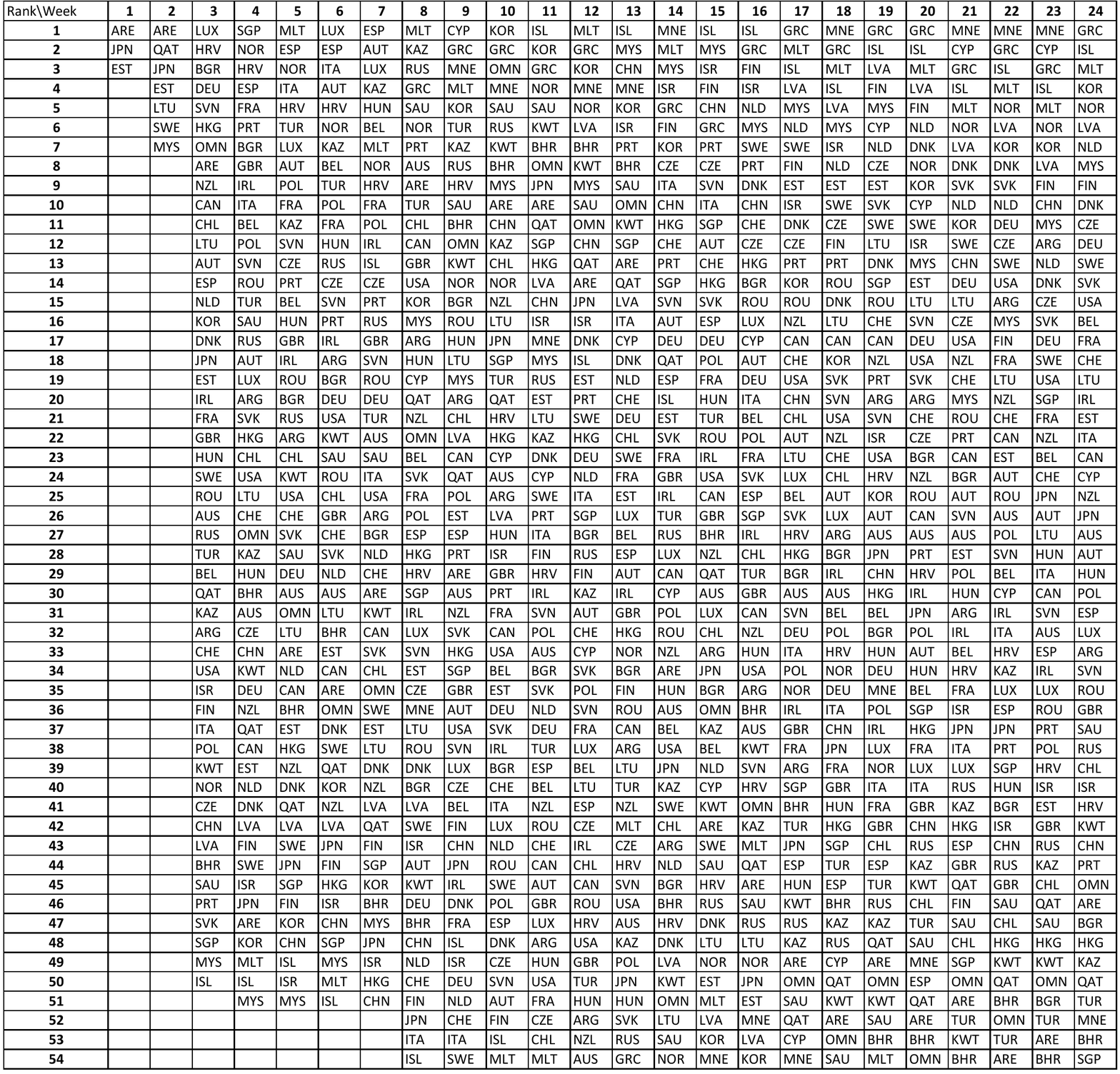}
\caption{The ranked data week by week. Columns represent weeks, while rows are ranks. Specifically, countries are sorted in descending order on the basis of the value of $A_j([t_1,t_2])$. The codes are taken from  ISO 3166-1, alpha-3.}
\label{tab:9}
\end{table}

\begin{table}[!htb]
\centering
\includegraphics[scale = 0.75]{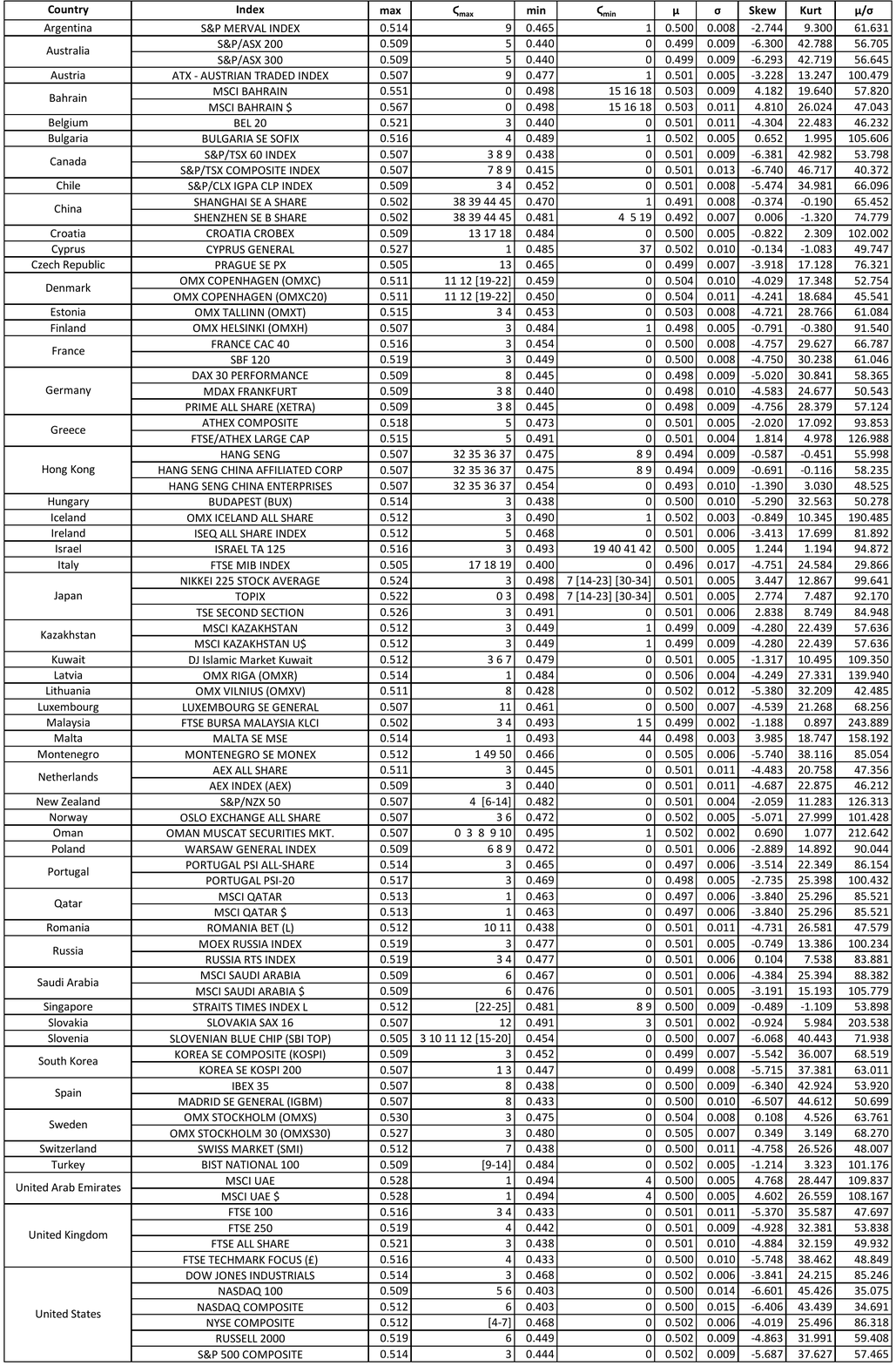}
\caption{Main statistical indicators of $H_j^{(\zeta)}(k)$ in Eq. \eqref{H_j}, at stock index level. The  values of the reference thresholds $\zeta$s are also shown.}
\label{tab:5}
\end{table}

\begin{table}[!htb]
\centering
\includegraphics[scale = 0.75]{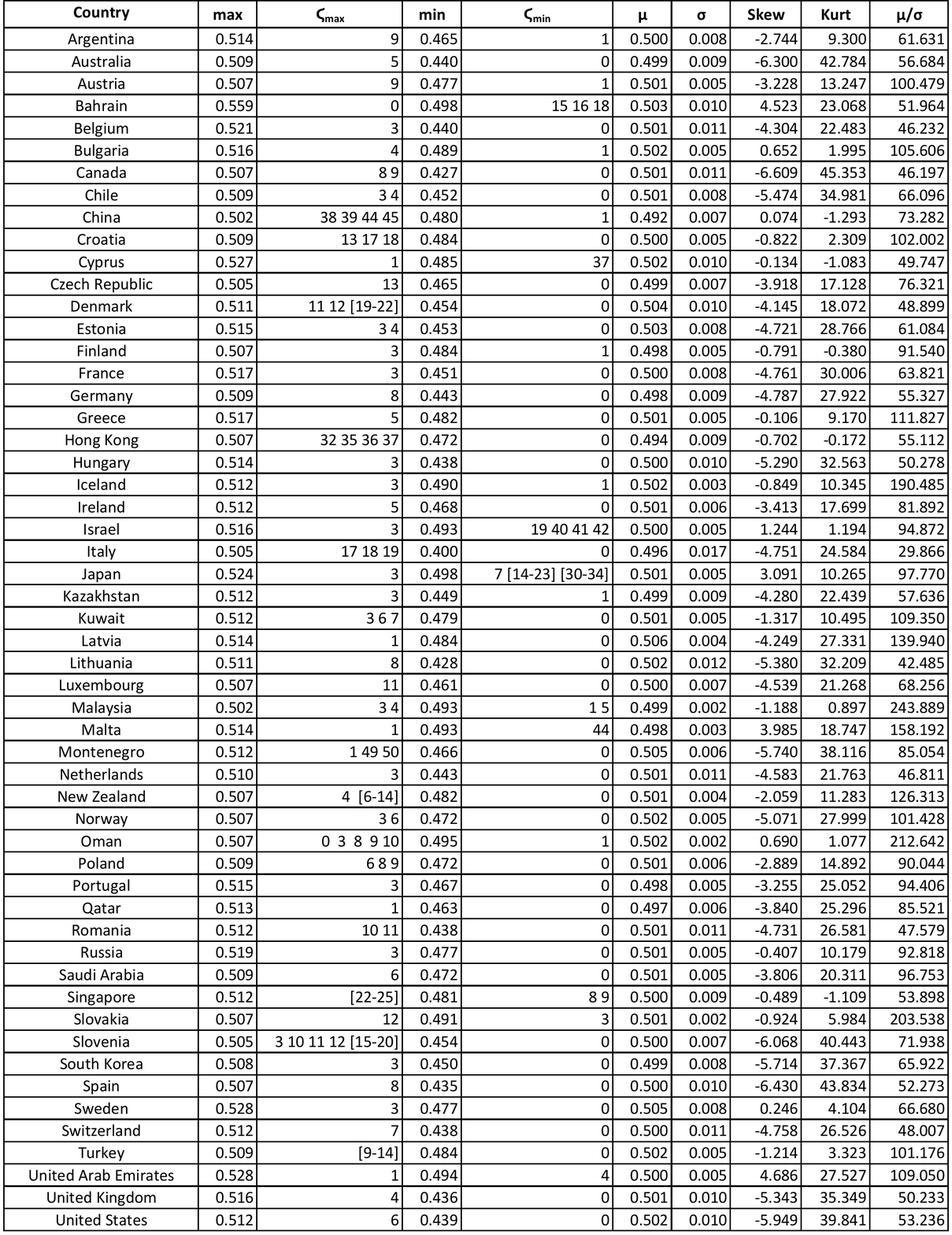}
\caption{Main statistical indicators of $H_j^{(\zeta)}$ in Eq. \eqref{H_jtot}, at country level. Also in this case, the  values of the reference thresholds $\zeta$s are illustrated.}
\label{tab:6}
\end{table}

\begin{table}[!htb]
\centering
\includegraphics[scale = 0.75]{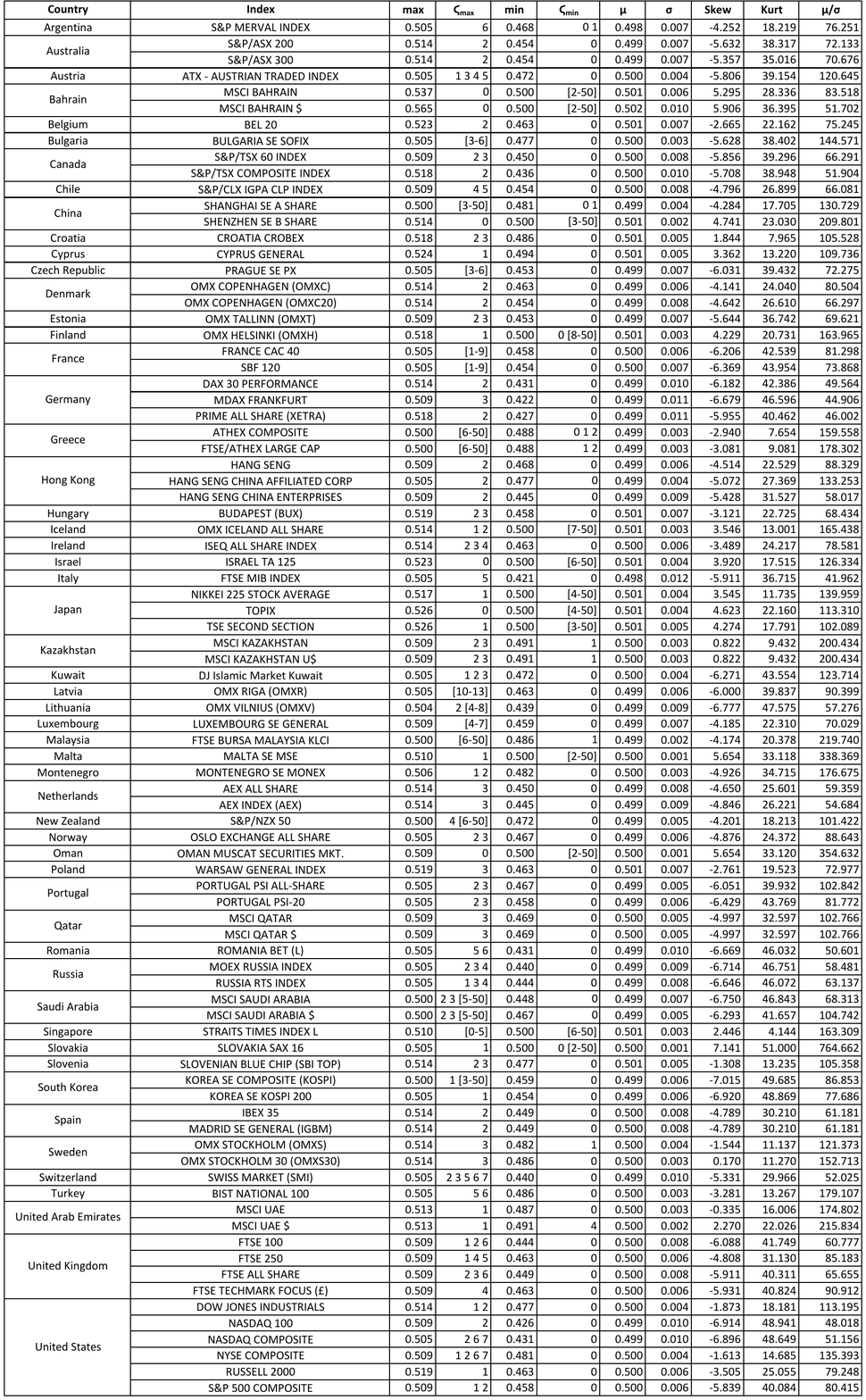}
\caption{Main statistical indicators of $R_j^{(\zeta)}(k)$ in Eq. \eqref{R_j}, at stock index level, along with the meaningful thresholds $\zeta$s.}
\label{tab:7}
\end{table}

\begin{table}[!htb]
\centering
\includegraphics[scale = 0.75]{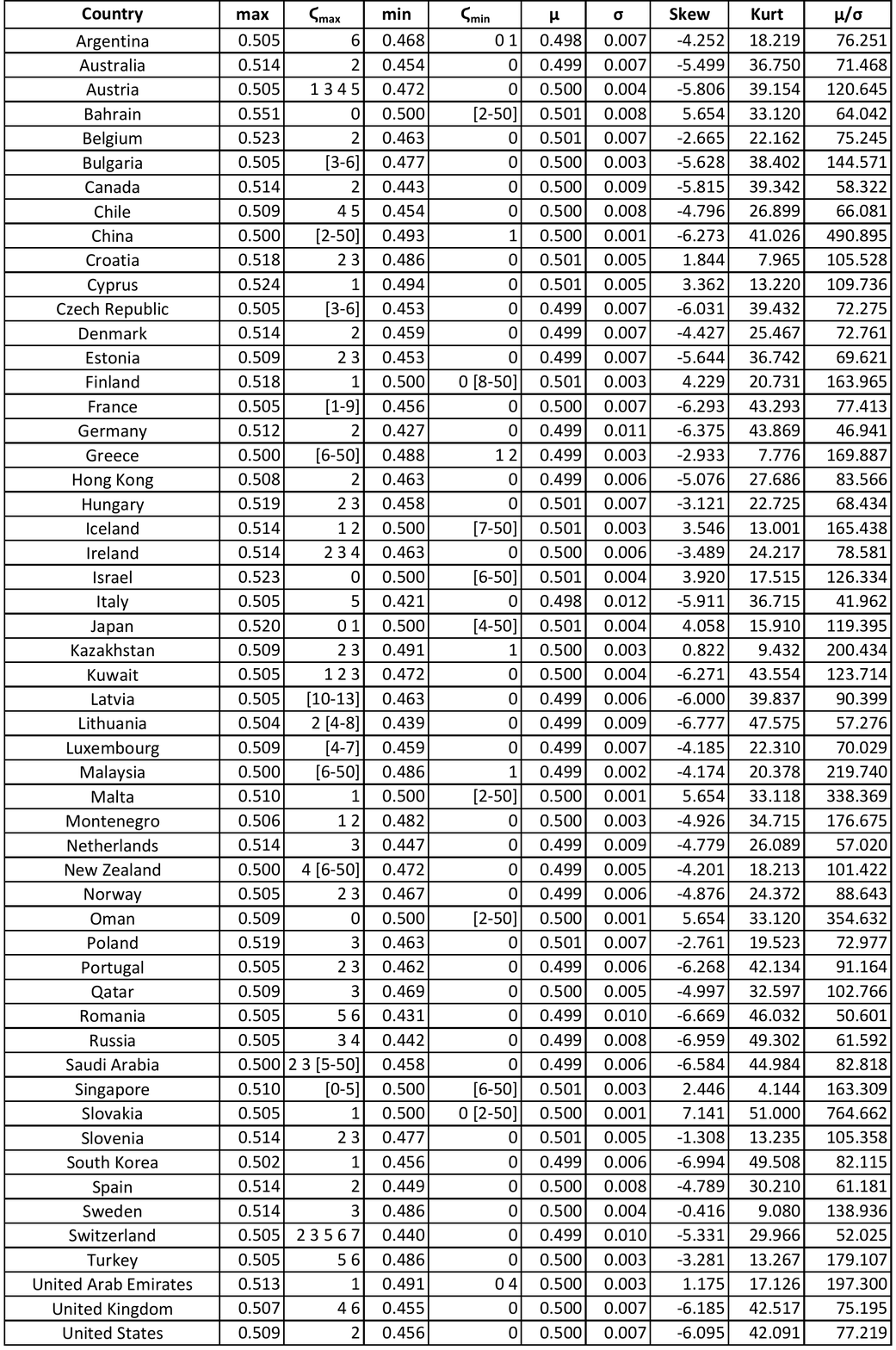}
\caption{Main statistical indicator of $R_j^{(\zeta)}$ in Eq. \eqref{R_jtot}, at country level. Also in this case, the reference thresholds $\zeta$s are reported.}
\label{tab:8}
\end{table}

\section{Conclusions}
\label{conclusioni}
The study investigates the relationship between the Google search volumes of \textit{``coronavirus''} and the stock index prices of different markets. The analysis is carried out at country level; thus, the word \textit{``coronavirus''} has been opportunely translated, when needed. Such analysis allows for mapping interrelationships between COVID-19 anxiety in nations and lack of trust in stock markets future performance. These aspects are related to the uncertainty surrounding the evolution of the pandemic and expectations about its effects. In our framework, we follow \cite{f20} and hypothesize that anxiety is manifested via the intensity of the searches run on Google related to the virus.%, while the prices are seen as information carriers, contain the manifestation of the disruptive event under analysis from a financial perspective.

The proposed indicators allow to capture changes in moods along the time -- for the case of the $A_j$'s in \eqref{A_j} and \eqref{A_jtot} -- and permit also classification of markets and countries under a more global perspective on the overall period -- see the $H_j$'s in \eqref{H_j} and \eqref{H_jtot} and the $R_j$'s in \eqref{R_j} and \eqref{R_jtot}. Moreover, the $A_j$'s compare the values of Google searches and prices, while the $H_j$'s and the $R_j$'s compare the daily increments/decrements of such quantities.

For a fair treatment of the considered dataset, we have taken only ``very high human developed countries'' -- i.e., those with an HDI greater than 0.8 -- and have reasonably added China. Some countries with HDI greater than 0.8 but without a stock exchange have been removed from the list.

The study allows having a panoramic view of the evolution of the pandemic, its effects on the behaviour of people and its impact on financial markets. Furthermore, the country-level approach gives insights on similarities and discrepancies of the different populations in respect of the link between the anxiety for COVID-19 and the expectations about stock markets performance.

%\begin{figure}[!htb]
%\includegraphics[width=\textwidth]{Bahraincase.png}
%\caption{The time series of the Google searches about ``coronavirus'' translation in Bahrain and the normalised MSCI Bahrain index.}
%\label{fig:10}
%\end{figure}
%
%
%\begin{figure}[!htb]
%\includegraphics[width=\textwidth]{Italiancase.png}
%\caption{The time series of the Google searches about ``coronavirus'' translation in Italy and the normalised FTSE MIB index.}
%\label{fig:11}
%\end{figure}


\clearpage

\begin{thebibliography}{10}

\bibitem{a90}
Robert~A Ariel.
\newblock High stock returns before holidays: Existence and evidence on
  possible causes.
\newblock {\em The Journal of Finance}, 45(5):1611--1626, 1990.

\bibitem{b06}
Robert~J Barro.
\newblock Rare disasters and asset markets in the twentieth century.
\newblock {\em The Quarterly Journal of Economics}, 121(3):823--866, 2006.

\bibitem{b20}
Carola Binder.
\newblock Coronavirus fears and macroeconomic expectations.
\newblock {\em The Review of Economics and Statistics}, 2020.
\newblock doi: 10.1162/rest\_a\_00931.

\bibitem{c13}
Yochi Cohen-Charash, Charles~A Scherbaum, John~D Kammeyer-Mueller, and Barry~M
  Staw.
\newblock Mood and the market: can press reports of investors' mood predict
  stock prices?
\newblock {\em PloS one}, 8(8):e72031, 2013.

\bibitem{colarossi2020}
Natalie Colarossi.
\newblock 8 times world leaders downplayed the coronavirus and put their
  countries at greater risk for infection.
\newblock
  \url{hhttps://www.businessinsider.com/times-world-leaders-downplayed-the-coronavirus-threat-2020-4?IR=T},
  2020.
\newblock [Online; accessed 19-June-2020].

\bibitem{f20}
Thiemo Fetzer, Lukas Hensel, Johannes Hermle, and Christopher Roth.
\newblock Coronavirus perceptions and economic anxiety.
\newblock {\em The Review of Economics and Statistics}, 2020.
\newblock doi: 10.1162/rest\_a\_00946.

\bibitem{ga12}
Xavier Gabaix.
\newblock Variable rare disasters: An exactly solved framework for ten puzzles
  in macro-finance.
\newblock {\em The Quarterly journal of economics}, 127(2):645--700, 2012.

\bibitem{ga20}
Dana~Rose Garfin, Roxane~Cohen Silver, and E~Alison Holman.
\newblock The novel coronavirus (covid-2019) outbreak: Amplification of public
  health consequences by media exposure.
\newblock {\em Health Psychology}, 39(5):355--357, 2020.

\bibitem{go90}
John~W Goodell.
\newblock Covid-19 and finance: Agendas for future research.
\newblock {\em Finance Research Letters}, 35:101512, 2020.

\bibitem{go12}
Francois Gourio.
\newblock Disaster risk and business cycles.
\newblock {\em The American Economic Review}, 102(6):2734--66, 2012.

\bibitem{ka00}
Mark~J Kamstra, Lisa~A Kramer, and Maurice~D Levi.
\newblock Losing sleep at the market: The daylight saving anomaly.
\newblock {\em The American Economic Review}, 90(4):1005--1011, 2000.

\bibitem{ka03}
Mark~J Kamstra, Lisa~A Kramer, and Maurice~D Levi.
\newblock Winter blues: A {SAD} stock market cycle.
\newblock {\em The American Economic Review}, 93(1):324--343, 2003.

\bibitem{ka10}
Guy Kaplanski and Haim Levy.
\newblock Sentiment and stock prices: {T}he case of aviation disasters.
\newblock {\em Journal of Financial Economics}, 95(2):174--201, 2010.

\bibitem{l20}
Qun Li, Xuhua Guan, Peng Wu, Xiaoye Wang, Lei Zhou, Yeqing Tong, Ruiqi Ren,
  Kathy~SM Leung, Eric~HY Lau, Jessica~Y Wong, et~al.
\newblock Early transmission dynamics in {W}uhan, {C}hina, of novel
  coronavirus--infected pneumonia.
\newblock {\em New {E}ngland {J}ournal of {M}edicine}, 382(13):1199--1207,
  2020.

\bibitem{m20}
Ga{\"e}tan Mertens, Lotte Gerritsen, Stefanie Duijndam, Elske Salemink, and
  Iris~M Engelhard.
\newblock Fear of the coronavirus ({COVID}--19): Predictors in an online study
  conducted in {M}arch 2020.
\newblock {\em Journal of Anxiety Disorders}, 74:102258, 2020.

\bibitem{KOR}
Dennis Normile.
\newblock Coronavirus cases have dropped sharply in south korea. what's the
  secret to its success?
\newblock
  \url{https://www.sciencemag.org/news/2020/03/coronavirus-cases-have-dropped-sharply-south-korea-whats-secret-its-success},
  2020.
\newblock [Online; accessed 21-July-2020].

\bibitem{ul1995reflections}
Mahbub Ul~Haq.
\newblock {\em Reflections on human development}.
\newblock Oxford University Press, 1995.

\bibitem{HDR2019}
UNDP.
\newblock {\em Human Development Report 2019}.
\newblock 2019.
\newblock doi: 10.18356/838f78fd-en.

\bibitem{world2020director}
WHO.
\newblock General's opening remarks at the media briefing on covid-19 -- 11
  march 2020.
\newblock
  \url{https://www.who.int/dg/speeches/detail/who-director-general-s-opening-remarks-at-the-media-briefing-on-covid-19---11-march-2020},
  2020.
\newblock [Online; accessed 1-June-2020].

\bibitem{wiki:ISL}
{Wikipedia contributors}.
\newblock Covid-19 pandemic in {I}celand --- {Wikipedia}{,} {T}he {F}ree
  {E}ncyclopedia.
\newblock
  \url{https://en.wikipedia.org/w/index.php?title=COVID-19_pandemic_in_Iceland&oldid=968774981},
  2020.
\newblock [Online; accessed 21-July-2020].

\bibitem{wiki:SWE}
{Wikipedia contributors}.
\newblock Covid-19 pandemic in {S}weden --- {Wikipedia}{,} {T}he {F}ree
  {E}ncyclopedia.
\newblock
  \url{https://en.wikipedia.org/w/index.php?title=COVID-19_pandemic_in_Sweden&oldid=968686163},
  2020.
\newblock [Online; accessed 21-July-2020].

\bibitem{z20}
Na~Zhu, Dingyu Zhang, Wenling Wang, Xingwang Li, Bo~Yang, Jingdong Song, Xiang
  Zhao, Baoying Huang, Weifeng Shi, Roujian Lu, et~al.
\newblock A novel coronavirus from patients with pneumonia in {C}hina, 2019.
\newblock {\em New England Journal of Medicine}, (382):727--733, 2020.

\end{thebibliography}
\end{document}